\documentclass[10pt,journal,compsoc]{IEEEtran}
\usepackage{amsmath,amssymb,amsfonts}
\usepackage{algorithmic}
\usepackage{graphicx}
\usepackage{textcomp}
\usepackage{xcolor}
\usepackage{amsthm}
\usepackage[ruled, linesnumbered, vlined]{algorithm2e}

\usepackage{pifont} 

\usepackage{textcomp}
\usepackage{multirow}
\usepackage{bm}
\usepackage{booktabs}
\usepackage{epstopdf}
\usepackage{cases}
\usepackage{makecell,multirow,diagbox}
\usepackage{stfloats}
\usepackage{enumitem}[topsep=3pt,itemsep=3pt]
\setlist{nolistsep}
\usepackage{comment}
\usepackage{xcolor}

\newtheorem{lemma}{Lemma}
\newtheorem{theorem}{Theorem}
\newtheorem{remark}{Remark}

\newenvironment{myproof}{\noindent\textit{Proof.}}{\hfill $\blacksquare$}

\usepackage{subfigure} 
\definecolor{color}{rgb}{0.0, 0, 0}
\definecolor{color1}{rgb}{0.0, 0, 0}
\definecolor{b2}{rgb}{0, 0, 0}
\definecolor{b3}{rgb}{0, 0, 0}

\usepackage{enumitem}[topsep=3pt,itemsep=3pt]
\allowdisplaybreaks[4]

\def\BibTeX{{\rm B\kern-.05em{\sc i\kern-.025em b}\kern-.08em
		T\kern-.1667em\lower.7ex\hbox{E}\kern-.125emX}}

\ifCLASSOPTIONcompsoc
\usepackage[nocompress]{cite}
\else
\usepackage{cite}
\fi

\begin{document}

\title{Multi-objective Optimization for Multi-UAV-assisted Mobile Edge Computing}

\author{Geng Sun,~\IEEEmembership{Member,~IEEE},
    	Yixian Wang,
		Zemin Sun,\IEEEmembership{}
        Qingqing Wu,~\IEEEmembership{Senior Member,~IEEE}, \\
        Jiawen Kang,~\IEEEmembership{Member,~IEEE},
        Dusit Niyato,~\IEEEmembership{Fellow,~IEEE}, and
        Victor C. M. Leung,~\IEEEmembership{Life Fellow,~IEEE}
  \thanks{This study is supported in part by the National Natural Science Foundation of China (62272194, 62172186), and in part by the Science and Technology Development Plan Project of Jilin Province (20230201087GX). (\textit{Corresponding author: Zemin Sun.)}}
	\IEEEcompsocitemizethanks{\IEEEcompsocthanksitem Geng Sun is with the College of Computer Science and Technology, Jilin University, Changchun 130012, China, and also with the Key Laboratory of Symbolic Computation and Knowledge Engineering of Ministry of Education, Jilin University, Changchun 130012, China. He is also with the College of Computing and Data Science, Nanyang Technological University, Singapore 639798.\protect\\
    E-mail: sungeng@jlu.edu.cn.
    \IEEEcompsocthanksitem Yixian Wang and Zemin Sun are with the College of Computer Science and Technology, Jilin University, Changchun 130012, China, and also with the Key Laboratory of Symbolic Computation and Knowledge Engineering of Ministry of Education, Jilin University, Changchun 130012, China.\protect\\
    E-mail: yixian23@mails.jlu.edu.cn, sunzemin@jlu.edu.cn.
    \IEEEcompsocthanksitem Qingqing Wu is with the Department of Electronic Engineering, Shanghai Jiao Tong University, Shanghai, China.\protect\\
    E-mail: qingqingwu@sjtu.edu.cn. 
    \IEEEcompsocthanksitem Jiawen Kang is with the School of Automation, Guangdong University of Technology, Guangzhou 510006, China.\protect\\
    E-mail:kjwx886@163.com.
    \IEEEcompsocthanksitem Dusit Niyato is with the College of Computing and Data Science, Nanyang Technological University, Singapore 639798.\protect\\
    E-mail: dniyato@ntu.edu.sg.
    \IEEEcompsocthanksitem Victor C. M. Leung is with the College of Computer Science and Software Engineering, Shenzhen University, Shenzhen 518060, China, and also with the Department of Electrical and Computer Engineering, University of British Columbia, Vancouver, BC V6T 1Z4, Canada. \protect\\
    E-mail: vleung@ieee.org.}
    \thanks{A small part of this paper is accepted by IEEE ICC 2024.}
}

\IEEEtitleabstractindextext{
\begin{abstract}	
Recent developments in unmanned aerial vehicles (UAVs) and mobile edge computing (MEC) have provided users with flexible and resilient computing services. However, meeting the computing-intensive and latency-sensitive demands of users poses a significant challenge due to the limited resources of UAVs. To address this challenge, we present a multi-objective optimization approach for multi-UAV-assisted MEC systems. First, we formulate a multi-objective optimization problem \textcolor{b2}{aiming} at minimizing the total task completion delay, reducing the total UAV energy consumption, and maximizing the total amount of offloaded tasks by jointly optimizing task offloading, computation resource allocation, and UAV trajectory control. Since the problem is a mixed-integer non-linear programming (MINLP) and NP-hard problem which is challenging, we propose a joint task offloading, computation resource allocation, and UAV trajectory control (JTORATC) approach to solve the problem. \textcolor{b3}{However, since the decision variables of task offloading, computation resource allocation, and UAV trajectory control are coupled with each other, the original problem is split into three sub-problems, i.e., task offloading, computation resource allocation, and UAV trajectory control, which are solved individually to obtain the corresponding decisions.} \textcolor{b2}{Moreover, the sub-problem of task offloading is solved by using distributed splitting and threshold rounding methods, the sub-problem of computation resource allocation is solved by adopting the Karush-Kuhn-Tucker (KKT) method, and the sub-problem of UAV trajectory control is solved by employing the successive convex approximation (SCA) method.} Simulation results show that the proposed JTORATC has superior performance compared to the other benchmark methods. 
\end{abstract}

}

\maketitle
\IEEEdisplaynontitleabstractindextext
\IEEEpeerreviewmaketitle

\section{Introduction}
\label{sec_introduction}
\par  \IEEEPARstart{W}{ith} the rapid development of 6G and the Internet of Things (IoT), the number of smart mobile devices has shown unprecedented growth, leading to the increasing proliferation of various innovative mobile applications \cite{Liu2019}. Most of these applications, such as face recognition, automatic navigation, and image processing, require intensive computation resources and low latency \cite{Asim2020}. However, handling the computation-hungry and real-time data generated by these applications is challenging due to the limited resources of mobile devices \cite{Abualigah2021}. In this context, mobile edge computing (MEC) has been regarded as a promising solution that allows mobile devices to offload the computation-intensive and delay-sensitive tasks to proximate edge servers \cite{Mach2017}, thereby reducing the computation burden, execution delay, and energy consumption of devices \cite{Porambage2018, Sun2024 ,Qu2022}. However, conventional MEC network relies on terrestrial infrastructures which are inflexible to deploy due to the installation cost and environmental limitations. 

\par To overcome the physical restrictions of traditional terrestrial MEC systems, unmanned aerial vehicles (UAVs)-assisted MEC is emerging to offer flexible and low-cost offloading services due to the high maneuverability, flexibility, fast deployment, and line-of-sight (LoS) links of UAVs \cite{Khawaja2019, Li2023, Ei2022,10382630}. By offloading the computation tasks to the proximate UAVs, mobile users can flexibly enjoy cloud computing services anywhere and anytime \cite{Qu2021,Dong2021,Qu2024}.

\par However, designing an efficient task offloading approach in the multi-UAV-assisted MEC systems still faces several challenges. \textcolor{b3}{The design of efficient strategy could face the following challenges.} \textit{First}, due to the heterogeneous and strict requirements of users, it is complex to determine efficient task offloading decisions that can satisfy the diverse requirements of different users under the resource constraints of UAVs. \textit{Second}, unlike the powerful cloud computing, the UAV-assisted MEC servers are equipped with restricted computing capabilities due to the weight limitation of the UAVs. Therefore, it is challenging to efficiently allocate the limited computing resources of UAVs to guarantee the delay sensitivity and computation intensity of the tasks. \textit{Finally}, UAVs are intrinsically limited by the finite onboard energy, which further limits the endurance of flight and service. However, performing the computation-intensive tasks could be time-consuming. Consequently, energy-efficient trajectory control is crucial but challenging for UAVs to provide satisfactory offloading service within the limited flight time. \textcolor{b3}{From the perspective of problem formulation, most current research tends to concentrate on  optimizing a single performance metric of the system, such as time delay or energy consumption while overlooking the multi-objective optimization.} \textcolor{b3}{In terms of algorithm design, the methods of machine learning such as reinforcement learning generally face the challenge of difficult convergence and long training periods, especially in large-scale scenarios.}

\par To address the abovementioned challenges, this work proposes a multi-objective optimization problem of task offloading, computation resource allocation, and UAV trajectory control for multi-UAV assisted MEC systems. The main contributions are summarized as follows:

\begin{itemize}
      \item We consider a multi-UAV-assisted MEC system. Based on this system, we formulate a multi-objective optimization problem aimed at minimizing the total task completion delay, reducing the total UAV energy consumption, and maximizing the total amount of offloaded tasks by jointly optimizing task offloading, computation resource allocation, and UAV trajectory control. Moreover, the optimization problem is proven to be a non-convex and mixed-integer non-linear programming (MINLP) problem.
	
	\item We propose a joint task offloading, computation resource allocation and UAV trajectory control (JTORATC) approach to solve the problem. \textcolor{b3}{However, the decision variables of task offloading, computation resource allocation, and UAV trajectory control are coupled with each other.} Thus, we first split the problem into three sub-problems, i.e., task offloading, computation resource allocation, and UAV trajectory control. Then, \textcolor{b2}{the sub-problem of task offloading is solved by using distributed splitting and threshold rounding methods, and the sub-problem of computation resource allocation is solved by adopting the Karush-Kuhn-Tucker (KKT) method. Besides, the sub-problem of UAV trajectory control is solved by employing the successive convex approximation (SCA) method.} Ultimately, the proposed JTORATC approach effectively optimizes the performance of the \textcolor{b2}{entire} system, reduces the complexity of the problem, and improves the solution efficiency.
	
	\item Simulations are conducted and the results show that the proposed JTORATC achieves superior performance in terms of objective function value, total task completion delay, and total UAV energy consumption compared to several benchmark schemes. Moreover, we find that the proposed algorithm has better scalability in the considered scenarios.
\end{itemize}

\par The rest of this study is organized as follows. Section \ref{sec_related work} reviews the related work. Section \ref{sec_model} introduces the proposed system model and problem formulation. Section \ref{sec_solution} presents the JTORATC approach. Moreover, the simulation results are given in section \ref{sec_results}. Finally, section \ref{sec_conclusion} summarizes the paper. 

%
%
\section{Related Work}
\label{sec_related work}

\par In this section, we review the research work that are related to MEC, UAV-assisted MEC, task offloading, resource allocation, and UAV trajectory control, as well as optimization approaches. Moreover, we summarize the differences between the related works and this work in Table \ref{comparison}.

\begin{table*}
    \setlength{\abovecaptionskip}{0pt}%
    \setlength{\belowcaptionskip}{-1em}%
    \caption{Comparison between Related Works and This Work}
    \label{comparison}
    \centering
    \setlength{\tabcolsep}{2.4pt} 
    \begin{tabular}{|c|c|c|c|c|c|c|c|c|c|}
        \hline 
        & \multicolumn{2}{c|}{Edge computing architecture} 
        & \multicolumn{3}{c|}{Optimization objective} 
        & \multicolumn{3}{c|}{Optimization variables}
        & \multicolumn{1}{c|}{Method}\\
        \hline 
        Reference  
        & \begin{tabular}{c} 
                UAV-assisted\\ MEC
          \end{tabular}
        & \begin{tabular}{c} 
                Multi-UAV-\\assisted\\MEC 
          \end{tabular}
        & \begin{tabular}{c} 
                Service\\delay
          \end{tabular}
        & \begin{tabular}{c} 
                Energy\\consumption
          \end{tabular}
          & \begin{tabular}{c} 
                Offloaded\\tasks
          \end{tabular}
          & \begin{tabular}{c} 
                Task\\offloading
          \end{tabular}
          & \begin{tabular}{c} 
                Resource \\allocation
          \end{tabular}
          & \begin{tabular}{c} 
                 UAV\\Trajectory \\ Control
          \end{tabular}
          & \begin{tabular}{c} 
                Distributed splitting,\\ Threshold rounding,\\SCA, KKT
          \end{tabular}\\
        \hline 
        \cite{Zhang2016}  & \ding{53} & \ding{53} & \ding{53} & \checkmark & \ding{53} & \checkmark & \ding{53} & \ding{53} & \ding{53}\\
        \hline 
        \cite{Mao2016} & \ding{53} & \ding{53} & \checkmark  & \ding{53} & \ding{53} & \checkmark & \ding{53} & \ding{53} & \ding{53}\\
        \hline 
        \cite{Kuang2019} & \ding{53} & \ding{53} & \checkmark & \checkmark & \ding{53} & \checkmark & \checkmark & \ding{53} & \ding{53}\\
        \hline
        \cite{Li2020} & \checkmark  & \ding{53}  & \ding{53} & \checkmark & \ding{53}  & \checkmark  & \ding{53} & \ding{53} & \ding{53}\\
        \hline
        \cite{Zhang2020a} & \checkmark & \ding{53} & \ding{53} & \checkmark & \ding{53} & \checkmark & \ding{53} & \checkmark & \ding{53}\\
        \hline
        \cite{Tun2022} & \checkmark & \checkmark & \checkmark  & \ding{53} & \ding{53} & \checkmark & \checkmark &\ding{53} & \ding{53}\\
        \hline
        \cite{Guo2024} & \checkmark & \checkmark & \checkmark  & \ding{53} & \ding{53} & \checkmark & \checkmark &\ding{53} & \ding{53}\\
        \hline
        \cite{Ei2022} & \checkmark & \checkmark & \ding{53} & \checkmark & \ding{53} & \checkmark & \checkmark &\ding{53} & \ding{53}\\
        \hline
        \cite{Yu2020} & \checkmark & \ding{53} & \checkmark & \checkmark & \ding{53} & \checkmark & \checkmark  & \ding{53} & \ding{53}\\
        \hline
        \cite{Chen2022} & \checkmark & \ding{53} & \ding{53} & \checkmark & \ding{53} & \checkmark & \checkmark & \ding{53} & \ding{53}\\
        \hline
        \cite{Ji2020} & \checkmark & \checkmark  & \ding{53} & \ding{53}  & \checkmark & \checkmark & \ding{53} & \checkmark & \ding{53}\\
        \hline
        \cite{Lee2024} & \checkmark & \checkmark & \ding{53} & \checkmark & \ding{53}  & \checkmark & \ding{53} & \checkmark & \ding{53}\\
        \hline
        \cite{10238831} & \checkmark & \checkmark & \checkmark & \checkmark & \ding{53} & \ding{53} & \checkmark & \checkmark & \ding{53}\\
        \hline
        \cite{Wu2022}  & \checkmark & \checkmark & \ding{53} & \ding{53} & \ding{53} & \checkmark & \ding{53} & \ding{53} & \ding{53}\\
        \hline
        \cite{Zhang2020c} & \checkmark & \checkmark & \ding{53} & \ding{53} & \ding{53} & \checkmark & \checkmark & \checkmark & \ding{53}\\
        \hline
        This work & \checkmark & \checkmark & \checkmark & \checkmark & \checkmark & \checkmark & \checkmark & \checkmark & \checkmark\\
        \hline
    \end{tabular}
\end{table*}

\par In recent years, MEC has become an area of intense interest. By offloading the latency-sensitive, compute-intensive and energy-intensive computing tasks to MEC servers, the quality of service (QoS) of mobile users can be significantly improved. For example, Zhang et al. \cite{Zhang2016} studied energy-efficient computation offloading (EECO) mechanisms for MEC in 5G heterogeneous networks. The optimization problem is formulated to minimize the energy consumption of the offloading system while considering the energy cost of both task computing and file transmission. Furthermore, Mao et al. \cite{Mao2016} considered MEC systems with energy harvesting devices and proposed an online task offloading algorithm which jointly determines the offload decision, CPU cycle frequency, and transmitted power. Besides, Kuang et al. \cite{Kuang2019} studied the joint optimization problem of partial offload scheduling and resource allocation in MEC systems with multiple independent tasks, with the aim of minimizing the weighted sum of execution delay and energy consumption while ensuring the transmission power constraints of the tasks. However, the abovementioned works studied task offloading and resource allocation in MEC with fixed edge infrastructures, which may not be applicable to the remote or disaster scenarios with inadequate infrastructures. 

\par To be more flexible in meeting the mobile needs of UAVs in specific environments and to handle dynamic tasks efficiently, there has been a growing research interest in task offloading, resource allocation, and trajectory optimization for UAV-assisted MEC systems. For example, Li et al. \cite{Li2020} addressed the problem of maximizing UAV energy efficiency through the joint optimization of UAV trajectory, user transmit power, and computational load distribution. Moreover, the authors in \cite{Zhang2020a} explored the problem of minimizing the total energy consumption by optimizing bit allocation, time slot scheduling, power allocation, and UAV trajectory. However, these studies mainly focus on single-server MEC systems, which may not be applicable to real-world scenarios, especially for large-scale scenarios with many users.

\par To overcome the challenges above, more research has focused on multi-UAV-assisted MEC environments. For example, Tun et al. \cite{Tun2022} proposed a multi-UAVs assisted cooperative MEC system integrated with an MEC-enabled terrestrial base station (BS) to minimize the total delay experienced by mobile users, studying the joint offloading decision as well as the allocation of communication and computing resources under the energy constraints of mobile users and UAVs. Guo et al. \cite{Guo2024} proposed a software defined network enhanced cooperative multiple UAV-enabled aerial computing (MUEAC) system to minimize the processing delay of separable tasks, and studied the joint task scheduling and computing resource allocation problem under task data dependence and UAV energy consumption constraints. Moreover, Ei et al. \cite{Ei2022} studied a multi-UAVs assisted two-stage MEC system, where UAVs provide computing and relaying services to mobile devices to minimize the energy consumption of mobile devices (MDs) and UAVs by jointly optimizing task offloading, communication as well as computing resource allocation. However, all the studies above only optimize task offloading and resource allocation, and the UAVs are either stationary or follow a certain trajectory, which may not be suitable for dynamic or cooperative UAV flight scenarios. 

\par To make full use of the flexible mobility of UAVs, some researches are devoted to integrating UAV trajectory control into UAVs-assisted MEC systems for optimizing the path planning of UAVs in dynamic environments and achieving more efficient task execution and resource allocation. For example, Yu et al.\cite{Yu2020} focused on minimizing the weighted sum of service latency and UAV energy consumption of all IoT devices by jointly optimizing UAV location, communication and computing resource allocation, and task splitting decisions. Moreover, Chen et al. \cite{Chen2022} aimed to minimize UAV energy consumption by jointly optimizing user offloading decisions, UAV positioning, and computing resource allocation. However, these studies mainly concentrated on UAV deployment while overlooking the trajectory optimization of UAVs, which could lead to inaccurate decision-making due to the lack of real-time control, especially in dynamic and complex environments. To address these issues, researchers have turned their focus to trajectory optimization in multi-UAV-assisted MEC systems. For example, Ji et al. \cite{Ji2020} proposed a new scheme that maximizes minimum throughput among users of UAV services by jointly optimizing cache placement, UAV trajectory, and transmission power within a finite time period. Furthermore, Lee et al. \cite{Lee2024} investigated a multi-UAV-mobile edge computing (UAV-MEC) network. By controlling the UAV-MEC trajectory and offloading ratio, the energy consumption, queue stability, and energy consumption of mobile devices (MDs) are jointly optimized. Besides, Park et al. \cite{10238831} considered the deployment and optimization of MEC-UAVs for subterahertz communication in remote areas so that minimizing the energy consumption and latency of MEC-UAVs and mobile users.

\par In addition, researchers have adopted various methods to solve the maximum or minimum optimization problem in UAV-assisted MEC systems. For example, Wu et al. \cite{Wu2022} studied joint computational offloading, UAV role and location selection problems in hierarchical multi-coalition UAV-assisted MEC networks, and formulated discrete Stackelberg games with multiple leaders and followers to capture hierarchical features and discrete optimization. Moreover, Zhang et al. \cite{Zhang2020c} proposed the problem of maximizing the computational efficiency in the multi-UAV assisted MEC system, and the joint optimization of user association, CPU cycle frequency, power and spectrum resource allocation and UAV trajectory scheduling was carried out and the iterative optimization algorithm of double-cycle structure was used to find the optimal solution. However, the abovementioned methods may have high computational complexity, especially when involving large-scale systems or complex dynamic environments.

\par The limitations of these previous studies are summarized as follows. First, these works primarily concentrated on optimizing one single performance of the system, such as delay and energy. Second, most of these studies only consider partial decision variables, e.g., task offloading and resource allocation or UAV trajectory and task offloading. Finally, most of the algorithms in these studies have high complexity. To address the limitations of these studies, we formulate a joint task offloading, computing resource allocation, and UAV trajectory control optimization problem to jointly minimize the total task completion delay and the total UAV energy consumption, and maximize the total amount of offloaded tasks. Moreover, \textcolor{b2}{we adopt the methods of distributed splitting, threshold rounding, KKT, and SCA to reduce the computational complexity.} 

%
%
\section{System Model and Problem Formulation}
\label{sec_model}

\par In this section, we first present the system model, followed by the mobility model, communication model and computation model. Then, we formulate the multi-objective optimization problem. In addition, the notations associated with the system model are listed in Table \ref{tab_notation} for the sake of readability. 

{\color{color}

	\begin{table*}[t]
	\vspace{0em}
	\setlength{\abovecaptionskip}{0pt}%
	\setlength{\belowcaptionskip}{0pt}%
	\caption{Summary of Notations}
	\label{tab_notation}
	\renewcommand*{\arraystretch}{1}
	\color{color}
	\begin{center}
		\begin{tabular}{|p{.183\textwidth}|p{.32\textwidth}||p{.1\textwidth}|p{.28\textwidth}|}
			\hline
			\textbf{Symbol}&\textbf{Description}&\textbf{Symbol}&\textbf{Description}\\
			\hline
				$U$ & The number of users& $M$ & The number of UAVs\\ 
			\hline
				$\mathcal{U}$ & The set of users& $\mathcal{M}$ & The set of UAVs \\
			\hline
				$H$ & UAV flight altitude & $B$ &Channel bandwidth\\
			\hline
				$d_{u, m}[n]$&The distance between UAV $m$ and user $u$ in time slot $n$ &$g_{u, m}[n]$&Channel power gain of UAV $m$ and user $u$ in time slot $n$\\
			\hline
				$R_{u,m}[n]$ &The data transmission rate between user $u$ and UAV $m$ in time slot $n$ & ${s}$ & The rotor stiffness\\
			\hline
				$p_u$ &Transmit power of user $u$&	${\sigma^2}$ & Noise power\\
			\hline
				$D_u[n]$ &The size of the task &$C_u$ &	The computation intensity of the task\\
			\hline
				$f_{u}$&The computing capability of user $u$& $f_{m,u}[n]$& The computing capability allocated by UAV $m$ to user $u$\\
			\hline
			  $T_\mathrm{local}[n] $& The service delay of user $u$ to offload its task to UAV $m$ in time slot $n$  &$T_\mathrm{off}[n]$&The service delay of user $u$ to process task locally in time slot $n$ \\
			\hline
				$T_{m,u}^{\mathrm{comp}}[n]$&Task processing delay & $T_{u,m}^{\mathrm{trans}}[n]$ &Task transmission delay\\
			\hline
			  ${P_0}$ &The blade profile power of UAV in hover state &$P_{ind}$ & The induced power of UAV in hover state\\
			\hline
				$U_{tip}$&The blade tip speed of the rotor blade & $v_0$ &The average rotor induced speed\\
			\hline
				$d_0$ & The fuselage resistance ratio& $A$ & The area swept by UAV blades\\
			\hline
				$E_m^{\mathrm{fly}}[n]$ & UAV $m$ flight energy consumption& $E_{m,u}^{\mathrm{comp}}[n]$& UAV $m$ calculate energy consumption\\
			\hline
                $V_{min}$ &The minimum speed of UAV $m$&
				$V_{max}$ &The maximum speed of UAV $m$\\
                \hline
				$\kappa_m$&The effective switching capacitance of UAV $m$& $\kappa_u$&Effective switching capacitance of user $u$\\
                \hline
		\end{tabular}
	\end{center}
\end{table*}     
}

%
%
\subsection{System Model}
\label{sec_system_model}

\begin{figure}[!hbt]
    \setlength{\abovecaptionskip}{0pt}
    \setlength{\belowcaptionskip}{0pt}
    \centering
    \includegraphics[width=3.5in]{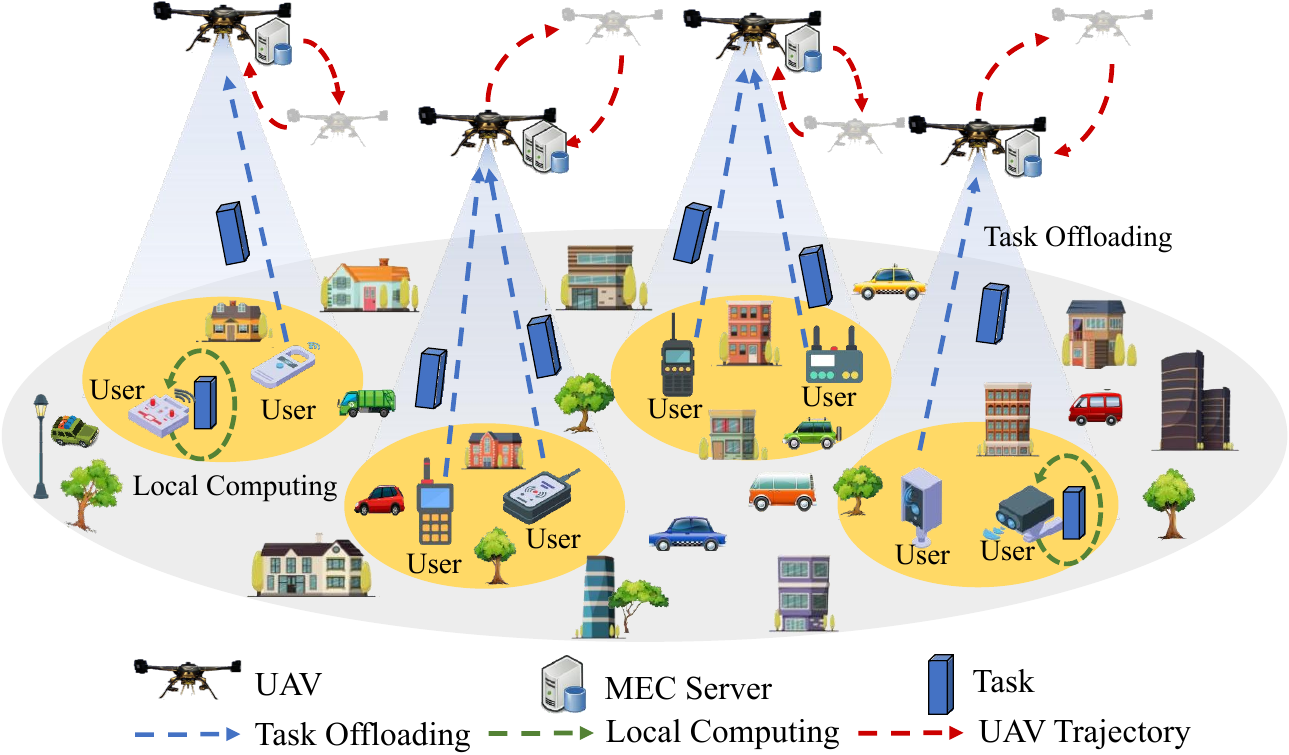}
    \caption{Multi-UAV-assisted MEC for task offloading.}
    \label{fig_gameModel}
\end{figure}

\par As shown in Fig. \ref{fig_gameModel}, a multi-UAV assisted MEC system is considered, which consists of $M$ rotary-wing UAVs and
$U$ users, denoted as $\mathcal{M}=\{1,2, \ldots,
M\}$ and $\mathcal{U}=\{1,2, \ldots, U\}$\textcolor{b2}{.} For ease of exposition, the continuous system timeline $T$ is discretized into $N$ time slots with equal time duration $\delta_t=T/N$, which is consistent with the coherence block of the wireless channel. Furthermore, \textcolor{b2}{each UAV is equipped with one or multiple MEC servers} to provide computing services to the users. Besides, in each time slot, each user could generate a task that can be executed locally or offloaded to a UAV. The task of user $u$ generated in time slot $n$ is characterized as $<D_u[n], C_u,\tau_u>$, wherein $D_u[n]$ is the task size (in bits), $C_u$ is computation intensity of the task (cycles/bit), and $\tau_u$ denotes the deadline of the task. 


\subsection{Mobility Model}
\label{sec_mobilityModel}

\par Without loss of generality, we consider a three-dimensional (3D) Cartesian coordinate system in which the horizontal position of each user $u\in\mathcal{U}$ is denoted as $\boldsymbol{w}_u=\left[x_u, y_u\right]^{\mathrm{T}}$. Moreover, we consider that each UAV flies at a constant altitude $H$ with the instantaneous horizontal position  $\mathbf{q}_m=\left[x_m[n], y_m[n]\right]^{\mathrm{T}}$, acceleration $a_m[n]$, and velocity $v_m [n]$ \cite{Wang2022a}. \textcolor{b2}{Let} $V_{\max}$ be the maximum speed of the UAV, then the maximum distance that each UAV can travel between the two stone troughs is expressed as $D_{\max}=V_{\max} \delta_t$. In addition, it is critical to ensure a safe distance \textcolor{b2}{among} UAVs to avoid collisions, which means that the distance between any two UAVs should be greater than the minimum safe distance in any time slot. Therefore, the mobility constraints of UAVs can be given as follows:
\begin{subequations}
	\label{eq_Mobility_constraint}
	\begin{alignat}{2}
          &\mathbf{q}_m[n+1]=\mathbf{q}_m[n]+v_m[n] \delta_t+\frac{1}{2} a_m[n] \delta_t^2,  \ \forall m, n, \label {1a}\\
         &v_m[n+1]=v_m[n]+a_m[n] \delta_t, \ \forall m, n,\label{1b}\\
           &\left\|v_m[n]\right\| \leq V_{\max }, \quad\left\|v_m[n]\right\| \geq V_{\min }, \  \forall m, n,\label {1c}\\
           &\left\|\mathbf{q}_m[n]-\mathbf{q}_i[n]\right\| \geq D_{\min }, \ \forall m \neq i, n,\label{1d}\\
            &\mathbf{q}_m[1]=\mathbf{q}_m[N], \ \forall m,\label{1e}\\
           &\left\|\mathbf{q}_m[n+1]-\mathbf{q}_m[n]\right\| \leq D_{\max }=V_{\max}\delta_t,  \ \forall m, n,\label{1f}
	\end{alignat}
\end{subequations}

\noindent where \eqref{1a} and \eqref{1b} \textcolor{b2}{denote} the position update and velocity update of UAVs, respectively, \eqref{1c} represents the speed constraints, \eqref{1d} ensures the safe distance between any two UAVs to avoid collisions, \eqref{1e} indicates that the initial and final positions of each UAV are fixed at the same location, and \eqref{1f} constrains the maximum flight distance during each time slot.

\subsection{Communication Model}
\label{sec_communicationModel}

\par The channel power gain between a user and a UAV is calculated by incorporating the probabilistic LoS transmissions into both small-scale and large-scale fading \cite{Yang2018}. For the uplink communication, the channel power gain between user $u$ and UAV $m$ at time slot $n$ can be given as:
\begin{equation}
\label{eq_channelPowerGain}
	g_{u,m}[n]=p_{u,m}^\mathrm{LoS}g_{u,m}^\mathrm{LoS}[n] +(1-p_{u,m}^\mathrm{LoS})g_{u,m}^\mathrm{NLoS}[n],
\end{equation}

\noindent where $p^\mathrm{LoS}$ represents the probability of LoS link, $g_{u,m}^\mathrm{LoS}[n]$ and $g_{u,m}^\mathrm{NLoS}[n]$ represent the channel power gain between user $u$ and UAV $m$ for LoS and NLoS links, respectively, which is calculated as:
{\color{color}
\begin{subequations}{\label{eq_channelPowerGain1}}
\begin{alignat}{2}
	&g_{u,m}^\mathrm{LoS}[n]=|h_{u,m}^\mathrm{LoS}[n]|^2\left(\mathcal{I}_{u,m}^\mathrm{LoS}[n]\right)^{-1}10^{\frac{-\mathcal{J}_{\sigma}^\mathrm{LoS}}{10}},\label{eq_channelPowerGainLoS1}\\
	&g_{u,m}^\mathrm{NLoS}[n]=|h_{u,m}^\mathrm{NLoS}[n]|^2\left(\mathcal{I}_{u,m}^\mathrm{NLoS}[n]\right)^{-1}10^{\frac{-\mathcal{J}_{\sigma}^\mathrm{NLoS}}{10}}, \label{eq_channelPowerGainNLoS1}
 \end{alignat}
\end{subequations}

\noindent where $h_{u,m}^\mathrm{LoS}[n]$ and $h_{u,m}^\mathrm{NLoS}[n]$, $\mathcal{I}_{u,m}^\mathrm{LoS}[n]$ and $\mathcal{I}_{u,m}^\mathrm{NLoS}[n]$, $\mathcal{J}_{\sigma}^\mathrm{LoS}$ and $\mathcal{J}_{\sigma}^\mathrm{NLoS}$ are the components of small-scale fading, path loss, and shadowing for LoS and NLoS links, respectively. For LoS link, these components are detailed as follows. 

\par First, the small-scale fading characteristic of the channel is captured by using a parametric-scalable and good-fitting generalized fading model, i.e., Nakagami-$m$ fading \cite{Boumaalif2022}, and to distinguish it from the identification of UAV, we use $w$ to represent the shape parameters in the Nakagami model, i.e., Nakagami-$w$. Specifically, $h_{u,m}^\mathrm{LoS}[n]$ and $h_{u,m}^\mathrm{NLoS}[n]$ follow \textcolor{b2}{the} Nakagami distribution with fading parameter $w^\mathrm{LoS}$ and $w^\mathrm{NLoS}$, which can be given as:
\begin{subequations}{\label{eq_LoS_probability}}
\begin{alignat}{2}
	&h_{u,m}^\mathrm{LoS}[n]=\frac{2{\left(w^\mathrm{LoS}\right)}^{w^\mathrm{LoS}}h_{u,m}^\mathrm{LoS}[n]^{2w^\mathrm{LoS}-1}e^{\left(-\frac{w^\mathrm{LoS}h^2}{\overline{p}}\right)}}{\Gamma\left(w^\mathrm{LoS}\right) \overline{p}^{w^\mathrm{LoS}}},\label{eq_channelPowerGainLoS3}\\
	&h_{u,m}^\mathrm{NLoS}[n]=\frac{2{\left(w^\mathrm{NLoS}\right)}^{w^\mathrm{NLoS}}h_{u,m}^\mathrm{NLoS}[n]^{2w^\mathrm{NLoS}-1}e^{\left(-\frac{w^\mathrm{NLoS}h^2}{\overline{p}}\right)}}{\Gamma\left(w^\mathrm{NLoS}\right) \overline{p}^{w^\mathrm{NLoS}}},\label{eq_channelPowerGainLoS4}
 \end{alignat}
\end{subequations}

\noindent \textcolor{b2}{where $\overline{p}$ is the average received power in the fading envelope}, and $\Gamma(w)$ is the Gamma function. 

\par Second, the path loss between user $u$ and UAV $m$ for LoS/NLoS link can be given as:
\begin{subequations}
\label{eq_LoS_probability1}
\begin{alignat}{2}
      &\mathcal{I}_{u,m}^\mathrm{LoS}[n]=\frac{\left(4\pi  d_0  f_c\right)^2}{c^2} \left(\frac{d_{u,m}[n]}{d_0}\right)^{\beta^\mathrm{LoS}},\label{eq_channelPowerGainLoS5}\\
      &\mathcal{I}_{u,m}^\mathrm{NLoS}[n]=\frac{\left(4\pi  d_0  f_c\right)^2}{c^2} \left(\frac{d_{u,m}[n]}{d_0}\right)^{\beta^\mathrm{NLoS}},\label{eq_channelPowerGainLoS6}
\end{alignat}
\end{subequations}
\noindent where $f_c$ is the carrier frequency, $c$ is the speed of light, $d_0$ is the reference distance,  $d_{u,m}[n]$ is the distance between user $u$ and UAV $m$, and $\beta^\mathrm{LoS}$ and $\beta^\mathrm{NLoS}$ are the path loss exponents for LoS and NLoS links, respectively. 

\par Finally, the shadowing captures the signal attenuation caused by shadowing in transmission, and it can be modeled as a zero-mean Gaussian distributed random variable, which is as follows:
\begin{subequations}
\label{eq_LoS_probability2}
\begin{alignat}{2}
      &\mathcal{J}_\sigma^\mathrm{LoS}[n]\sim\mathcal{O}\left(0,\left(\sigma^\mathrm{LoS}\right)^2\right),\label{eq_channelPowerGainLoS7}\\
      &\mathcal{J}_\sigma^\mathrm{NLoS}[n]\sim\mathcal{O}\left(0,\left(\sigma^\mathrm{NLoS}\right)^2\right),\label{eq_channelPowerGainLoS8}
\end{alignat}
\end{subequations}
\noindent where $\sigma^\mathrm{LoS}$ and $\sigma^\mathrm{NLoS}$ are the standard deviation of shadowing for LoS and NLoS links, respectively \cite{Yang2018}. 
}

\par Accordingly, in time slot $n$, the data transmission rate between user $u$ and UAV $m$ can be given as:
\begin{equation}
\label{eq_Transmission_rate}
	R_{u, m}[n]=B \log _2\big(1+\frac{p_u g_{u, m}[n]}{\sigma^2}\big),
\end{equation}

\noindent where $B$ is the channel bandwidth, $p_u$ is the transmit power of user $u$, and ${\sigma^2}$ is the noise power.

\subsection{Computation Model}
\label{sec_ComputationModel}

\par This section presents the service delay, energy consumption, and total amount of offloaded tasks.

\subsubsection{Service Delay} 

\par The service delay depends on the task offloading decision $a_{u,m}[n]$, which indicates whether the task of user $u$ is offloaded to UAV $m$ (i.e., $a_{u,m}[n]=1$) or processed locally (i.e., $a_{u,m}[n]=0$).

\textbf{Local Computing.} The service delay of user $u$ to process task $K_u[n]$ locally in time slot $n$ can be given as:
\begin{equation}
\label{eq_time_local}
	T_{\mathrm{local}}[n]=\frac{D_u[n]C_u}{f_{u}},
\end{equation}

\noindent where $f_{u}$ denotes the computing capability of user $u$.

\textbf{UAV-Assisted Computing.} The service delay of user $u$ to offload task $K_u[n]$ to UAV $m$ in time slot $n$ mainly consists of the transmission delay and processing delay. Specifically, the transmission delay can be given as:
\begin{equation}
\label{eq_time_transmission}
	T_{u,m}^{\mathrm{trans}}[n]=\frac{D_u[n]}{R_{u, m}[n]}.
\end{equation}

\noindent Moreover, the processing delay can be expressed as:
\begin{equation}
\label{eq_time_calculation}
	T_{m,u}^{\mathrm{comp}}[n]=\frac{D_u[n]C_u}{f_{m,u}[n]},
\end{equation}
\noindent where $f_{m,u}$ indicates the computing capability allocated by UAV $m$ to the user $u$.

\par Therefore, by combining Eqs. \eqref{eq_time_transmission} and \eqref{eq_time_calculation}, the service delay for UAV-assisted computing can be calculated as:
\begin{equation}
\label{eq_time_off}
	T_{\mathrm{off}}[n]=\frac{D_u[n]C_u}{f_{m,u}[n]}+\frac{D_u[n]}{R_{u, m}[n]}.
\end{equation}

\par According to Eqs. \eqref{eq_time_local} and \eqref{eq_time_off}, the total task completion delay for all users across $N$ time slots can be calculated as:
\begin{align}
     T_{\mathrm{total}}
     & = \sum_{n=1}^N \sum_{u=1}^U \sum_{m=1}^M a_{u,m}[n]T_{\mathrm{off}}[n] + (1 - a_{u,m}[n]) T_{\mathrm{local}}[n]\nonumber\\ 
     & =  \sum_{n=1}^N \sum_{u=1}^U \sum_{m=1}^M a_{u,m}[n] \left( T_{\mathrm{off}}[n] - T_{\mathrm{local}}[n] \right)+T_{local}[n].
\end{align}

\par Given that the results are typically much smaller in comparison to the input data for most applications, the result download delay is ignored \cite{Wang2022b}.

\subsubsection{Energy Consumption}
\label{sec_energyModel}

\par \textcolor{b2}{Completing tasks} could impose additional costs of energy consumption on both users and UAVs.

\par \textbf{\textit{Local Computing.}} The energy consumption of user $u$ to process task $K_u[n]$ locally in time slot $n$ can be given as:
\begin{equation}
\label{eq_energyLocal}
        E_{\mathrm{local}}^{\mathrm{comp}}[n]=\kappa_u(f_u)^2D_u[n]C_u,
\end{equation}
\noindent where $\kappa_u$ denotes the effective capacitance coefficient of user $u$, which depends on the chip structure of the CPU \cite{Burd1996}.	

\par \textbf{\textit{UAV-Assisted Edge Computing.}} The energy consumed by UAV $m$ to process task $K_u[n]$ in time slot $n$ can be given as \cite{Wang2016}:
\begin{equation}
\label{eq_energyMec}
	E_{m,u}^{\mathrm{comp}}[n]=\kappa_m(f_{m,u}[n])^2D_u[n]C_u,
\end{equation}
\noindent where $\kappa_m$ represents the effective switching capacitance that depends on the CPU architecture.

\par \textbf{\textit{UAV Fight.}} The propulsion power consumption of UAV $m$ flying at speed $v_m[n]$ can be given as\cite{Zeng2019,Zeng2019a}:
\begin{equation}
\label{eq_UAV_fight}
     \begin{aligned}
     E_m^{\mathrm{fly}}[n] = & \big(P_0\big(1+\frac{3\big\|v_m[n]\big\|^2}{U_{\text{tip}}^2}\big)+P_{\text{ind}}\big(\sqrt{1+\frac{\big\|v_m[n]\big\|^4}{4v_0^4}}\\
     &-\frac{\big\|v_m[n]^2\big\|}{2v_0^2}\big)^{1/2} + \frac{1}{2}d_0\rho_0|sA{\big\|v_m[n]^3\big\|}\big)\delta_t,
     \end{aligned}
\end{equation}

\noindent where ${P_0}$ is the blade profile power, $P_{ind}$ is the induced power, $U_{tip}$ is the blade tip speed of the rotor blade, and $v_0$ is the average rotor induced speed. Moreover, $d_0$ indicates the fuselage resistance ratio, $\rho_0$ is the air density, $s$ is the rotor stiffness, and $A$ indicates the area swept by UAV blades.

\par According to Eqs. \eqref{eq_energyMec} and \eqref{eq_UAV_fight}, the total energy consumption of the UAVs across $N$ time slots is mainly incurred by task execution and flight, which can be given as:
\begin{equation}
\label{eq_energyUAVfight}
E_{\mathrm{total}}=\sum_{n=1}^N\big(\sum_{u=1}^U\sum_{m=1}^M\big(E_m^{\mathrm{fly}}[n]+a_{u,m}[n]E_{m,u}^{\mathrm{comp}}[n]\big)\big).
\end{equation}

\subsubsection{Total Amount of Offloaded Tasks}
\label{sec_task}

\par The total amount of offloaded tasks can reflect the service quality provided by UAVs, which can be calculated as:
\begin{equation}
\label{eq_Total offload tasks}  
      K_{\mathrm{total}}=\sum_{n=1}^N \big(\sum_{u=1}^U\sum_{m=1}^Ma_{u,m}[n]D_u[n]\big).
\end{equation}

\subsection{Problem Formulation}
\label{sec_Problem Formulation}

\par This work aims to 
simultaneously minimize the total task completion delay, minimize the total UAV energy consumption, and maximize the total amount of offloaded tasks by optimizing the decisions of task offloading $\mathbf{A}=\left\{a_{u, m}[n], \forall u, m, n\right\}$, computation resource allocation $\mathbf{F}=\left\{f_{m,u}[n], \forall u \in \mathbf{U}_0, m, n\right\}$, and UAV trajectory control $\mathbf{Q}=\left\{\mathbf{q}_m[n], \forall m, n\right\}$. Accordingly, the multi-objective optimization problem can be formulated as:
\begin{subequations}
	\label{eq_problem}
	\begin{alignat}{2}
		\mathbf{P}: \quad & \min _{\mathbf{A},\mathbf{F},\mathbf{Q}}\quad\{T_{\mathrm{total}},\ E_{\mathrm{total}},\ -K_{\mathrm{total}}\}\label{12a}\\
		\text{s.t.} \quad 
             &a_{u,m}[n]\in \{0,1\}, \ \forall u,m,n,\label{4a}\\
             &\sum_{m=1}^M a_{u,m}[n]\leq1, \ \forall u,m,n,\label{4b}\\
             &\sum_{u=1}^U a_{u,m}[n] \leq 1, \ \forall u,m,n,\label{4c}\\
             &a_{u,m}[n]\cdot T_{\mathrm{local}}[n]\leq \tau_u,\ \forall u,m,n,\label{12d}\\
             &a_{u,m}[n]\cdot T_{\mathrm{off}}[n]\leq \tau_u,  \ \forall u,m,n,\label{12e}\\
             &0 \leq f_{m,u}[n] \leq f_m^{\text{max}}, \ \forall m,n,\label{12g}\\
             &\sum_{u=1}^\mathbf{U_0} f_{m,u}[n] \leq f_m^{\text{max}},  \ \forall u,m,n, \label{12h}\\
             & \eqref{1a} \sim \eqref{1f}. \label{12f}
	\end{alignat}
\end{subequations}

\noindent Constraints \eqref{4a}-\eqref{4c} represent the task offloading constraints of users. Constraints \eqref{12d} and \eqref{12e} indicate the deadline of the tasks. Constraints \eqref{12g} and \eqref{12h} limit the computation resource of UAV. Moreover, constraint \eqref{12f} represents the mobility constraint of UAV. 
\begin{theorem}
	\label{lemma_NP}
	{Problem $\mathbf{P}$ is a non-convex and NP-hard MINLP.}
\end{theorem}

\begin{myproof}
        Problem $\mathbf{P}$ involves binary variables (i.e., task offloading $\mathbf{A}$) and continuous variables (i.e., computation resource allocation $\mathbf{F}$ and UAV trajectory control $\mathbf{Q}$), while the inequalities in \eqref{1c} and \eqref{1d} are non-convex constraints. Consequently, problem $\mathbf{P}$ is \textcolor{b2}{an} MINLP problem, which is also non-convex and NP-hard. 
\end{myproof}

\par It can be deduced from Theorem \ref{lemma_NP} that it is difficult to find an optimal solution to problem $\mathbf{P}$ due to the NP hardness, which motivates us to propose the JTORATC. 

%
%
\section{The Proposed JTORATC}
\label{sec_solution}

\par In this section, the JTORATC is proposed to solve the formulated optimization problem. First, we transform the original multi-objective optimization problem into a single-objective optimization problem to simplify complexity, unify evaluation criteria, and facilitate trade-offs of different objectives. Then, we split the single-objective optimization problem into three sub-problems of task offloading, computation resource allocation, and UAV trajectory control based on the block alternate descent method, which alternates the solution of each sub-problem iteratively. \textcolor{b2}{Specifically, we employ distributed splitting and threshold rounding methods to solve the sub-problem of task offloading. Moreover, the sub-problem of computation resource allocation is solved by adopting KKT method, and the sub-problem of UAV trajectory control is solved by employing SCA method.}

\subsection{Problem \textcolor{b2}{Reformulation}}
\label{sec_Problem Reformulation}

\par Similar to \cite{Alsyouf2017}, we convert the multi-objective optimization problem into a single-objective optimization problem by using the right-of-use coefficient method. Consequently, the objective function can be rewritten as:
\begin{equation}
\begin{aligned}
&\rho(\mathbf{A}, \mathbf{F}, \mathbf{Q}) 
=  w_1T_{\mathrm{total}} + w_2E_{\mathrm{total}} - w_3K_{\mathrm{total}}\\
&= \sum_{n=1}^N\sum_{u=1}^U\sum_{m=1}^M 
     w_1\left(a_{u,m}[n] \left( T_{\mathrm{off}}[n] - T_{\mathrm{local}}[n] \right)
    +T_{local}[n]\right)\\
    &+w_2\left(E_m^{\text{fly}}[n]+a_{u,m}[n]E_{m,u}^{\text{comp}}[n]\right)-w_3\left(a_{u,m}[n]D_u[n]\right),
\end{aligned}
\end{equation}

\noindent where the non-negative parameters $w_1$, $w_2$, and $w_3$ denote the weight factors of the delay, energy consumption, and amount of offloaded tasks, respectively. Note that the weight coefficients can be adjusted according to the user preferences on the objectives. Then, the original multi-objective optimization problem $\mathbf{P}$ is transformed into a single-objective optimization problem $\mathbf{P_1}$, which is expressed as follows:
\begin{subequations}
	\label{eq_problem1}
	\begin{alignat}{2}
		\mathbf{P_1}: \quad & \min _{\mathbf{A},\mathbf{F}, \mathbf{Q}}\quad\rho\left({\mathbf{A}},{\mathbf{F},\mathbf{Q}}\right)\label{15a}\\
		\text{s.t.} \quad  &a_{u,m}[n]\in \{0,1\}, \ \forall u,m,n,\label{4.1a}\\
             &\sum_{m=1}^M a_{u,m}[n]\leq1, \ \forall u,m,n,\label{4.1b}\\
             &\sum_{u=1}^U a_{u,m}[n] \leq 1, \ \forall u,m,n,\label{4.1c}\\
             &a_{u,m}[n]\cdot T_{\mathrm{local}}[n]\leq \tau_u,\ \forall u,m,n,\label{4.1d}\\
             &a_{u,m}[n]\cdot T_{\mathrm{off}}[n]\leq \tau_u,  \ \forall u,m,n,\label{4.1e}\\
             &0 \leq f_{m,u}[n] \leq f_m^{\text{max}}, \ \forall m,n,\label{4.1f}\\
             &\sum_{u=1}^\mathbf{U_0} f_{m,u}[n] \leq f_m^{\text{max}},  \ \forall u,m,n, \label{4.1g}\\
             & \eqref{1a} \sim \eqref{1f}. \label{4.1h}
	\end{alignat}
\end{subequations}

\par This transformation reduces the difficulty of the original formulated optimization problem, avoids the metric differences between different objectives, and permits flexible handling of objectives through weight adjustments.

{\color{b3}\par However, the transformed optimization problem is still difficult to solve due to the convex objective function and non-convex constraints. Considering that the decision variables of task offloading, computation resource allocation, and UAV trajectory control are coupled with each other, we split the single-objective optimization problem into three sub-problems of task offloading, computation resource allocation, and UAV trajectory control based on the block alternate descent method.
\begin{remark}
	\label{remark1}
        If problem $\mathbf{P_1}$ is split into three sub-problems, each sub-problem can be transformed into a convex optimization problem and can be solved using the existing methods. However, splitting problem $\mathbf{P_1}$ into two sub-problems may compromise flexibility, solving efficiency, and complexity. For example, if the original problem is split into the sub-problems of $\mathbf{P_{1.1}}: \min_{\mathbf{A},\mathbf{F}}\rho\left(\mathbf{A},\mathbf{F}\right)$ and $\mathbf{P_{1.2}}: \min_{\mathbf{Q}}\rho\left(\mathbf{Q}\right)$, or into the sub-problems of $\mathbf{P_{1.1}}: \min_{\mathbf{A},\mathbf{Q}}\rho\left(\mathbf{A},\mathbf{Q}\right)$ and $\mathbf{P_{1.2}}: \min_{\mathbf{F}}\rho\left(\mathbf{Q}\right)$, combining  binary variable $\mathbf{A}$ and continuous variable $\mathbf{F}$ or $\mathbf{Q}$ into a single sub-problem might limit the ability to apply different methods tailored to each variable type. Furthermore, if the problem is split into sub-problems of $\mathbf{P_{1.1}}: \min_{\mathbf{A}}\rho\left(\mathbf{A}\right)$ and $\mathbf{P_{1.2}}: \min_{\mathbf{F},\mathbf{Q}}\rho\left(\mathbf{F},\mathbf{Q}\right)$, the mixed continuous variables could further add the complexity on solving the problem. Thus, it is reasonable to split the original optimization problem into three sub-problems.
        
 \end{remark}
}
\subsection{Task Offloading}
\label{sec_Task Offloading}

\par Given the computation resource allocation $\widehat{\mathbf{F}}=\{\widehat{\mathbf{F}}_{m,u}[n], \forall u \in \mathbf{U}_0,m,n\}$ and UAV trajectory control $\widehat{\mathbf{Q}}=\{\widehat{\mathbf{q}}_m[n], \forall m,n\}$,  the task offloading optimization problem can be expressed as:
\begin{subequations}
	\label{eq_Task offloading}
	\begin{alignat}{2}
		\mathbf{P_{1.1}}: \quad & \min _{\mathbf{A}}\quad\rho\left({\mathbf{A}},\widehat{\mathbf{F}},\widehat{\mathbf{Q}}\right)\label{18a}\\
		\text{s.t.}  \quad 
             &a_{u,m}[n]\in \{0,1\}, \ \forall u,m,n,\label{1.1b}\\
             &\sum_{m=1}^M a_{u,m}[n]\leq1, \ \forall u,m,n,\label{1.1c}\\
             &\sum_{u=1}^U a_{u,m}[n] \leq 1, \ \forall u,m,n,\label{1.1d}\\
             &a_{u,m}[n]\cdot T_{\mathrm{local}}[n]\leq \tau_u,\ \forall u,m,n,\label{1.1e}\\
             &a_{u,m}[n]\cdot T_{\mathrm{off}}[n]\leq \tau_u,  \ \forall u,m,n.\label{1.1f}	
	\end{alignat}
\end{subequations}

{\color{b2}
\par Problem $\mathbf{P_{1.1}}$ is still non-convex due to the constraint \eqref{1.1b}, which involves the binary variable $a_{u,m}[n]$. Therefore, we first transform problem $\mathbf{P_{1.1}}$ into a convex problem by using the distributed splitting method. Then, the threshold rounding method is adopted to convert the relaxed continuous variables back into binary variables.


\par For the non-convex constraint \eqref{1.1b}, we first employ the distributed splitting method to transform the binary variables into continuous variables. In specific, the variable of task offloading $\mathbf{A}$
is split into two parts, i.e., $\mathbf{X}$ and $\mathbf{Y}$ as follows:
\begin{subequations}
    \begin{alignat}{2}
     \mathbf{X}= & \left\{x_a^u[n] \mid x_a^u[n]=1-a^u[n], \ \forall u, n\right\}, \\
     \mathbf{Y}= & \left\{y_a^m[n] \mid y_a^m[n]=a^m[n], \ \forall m, n\right\},
    \end{alignat}
\end{subequations}
\noindent where $a^u[n]=1$ indicates that the task is processed locally, and $a^m[n]=1$ means being offloaded to the UAV. Then, We slack the variables into continuous variables as follows:
\begin{subequations}
   \begin{alignat}{2}
    & \overline{\mathbf{X}} \triangleq\left\{\sum_{u=1}^U x_a^u[n]=1, x_a^u[n] \in[0,1]\right\}, \label{X}\\
    & \overline{\mathbf{Y}} \triangleq\left\{\sum_{m=1}^M y_a^m[n]=1, y_a^m[n] \in[0,1]\right\}.\label{Y}
    \end{alignat}
\end{subequations}
\noindent Based on the above steps, problem $\mathbf{P_{1.1}}$ is transformed into problem $\mathbf{P_{1.1^{\prime}}}$ as follows:
\begin{subequations}
	\label{eq_Task offloading1.11}
	\begin{alignat}{2}
		\mathbf{P_{1.1^{\prime}}}: \quad & \min _{\overline{\mathbf{X}},\overline{\mathbf{Y}}}\quad\rho\left({\overline{\mathbf{X}},\overline{\mathbf{Y}}},\widehat{\mathbf{F}},\widehat{\mathbf{Q}}\right)\label{1.11a}\\
		\text{s.t.}  \quad 
             &\sum_{m=1}^M y_a^m[n]\leq1, \ \forall u,m,n,\label{1.11c}\\
             &\sum_{u=1}^U x_a^u[n] \leq 1, \ \forall u,m,n,\label{1.11d}\\
             &x_a^u[n]\cdot T_{\mathrm{local}}[n]\leq \tau_u,\ \forall u,m,n,\label{1.11e}\\
             &y_a^m[n]\cdot T_{\mathrm{off}}[n]\leq \tau_u,  \ \forall u,m,n,\label{1.11f}\\
             &\eqref{X},\eqref{Y}. \label{1.11g}
	\end{alignat}
\end{subequations}

\begin{theorem}
	\label{the_opt_convex1}
        \textcolor{b2}{Problem $\mathbf{P_{1.1^{\prime}}}$ is a convex non-linear programming (NLP) problem.}
\end{theorem}

\begin{myproof} Since the computation resource allocation and UAV trajectory control have been fixed, $E_m^{\text{fly}}[n]$ and $E_{m,u}^{\text{comp}}[n]$ are fixed values that do not affect the convexity of problem $\mathbf{P_{1.1}^{\prime}}$.  Consequently, the objective function $\rho$ is the sum of three linear expressions about $x_a^u[n]$ and $y_a^m[n]$, rendering it a linear function. Besides, two variables (i.e., $\overline{\mathbf{X}}$ and $\overline{\mathbf{Y}}$) are involved in the sub-problem. Therefore, problem $\mathbf{P_{1.1^{\prime}}}$ is a convex NLP problem\cite{Zhang2020}.
\end{myproof}

\par According to Theorem \ref{the_opt_convex1}, solving problem $\mathbf{P_{1.1^{\prime}}}$ directly remains challenging since it is a convex NLP. Therefore, we
iteratively optimize $\mathbf{\overline{X}}$ and $\mathbf{\overline{Y}}$ by splitting $\mathbf{P_{1.1^{\prime}}}$. Specifically, it can be expressed as:
\begin{subequations}
	\label{eq_Task offloading1.11_X}
	\begin{alignat}{2}
		\mathbf{P_{1.1^{\prime}}^{\overline{X}}}: \quad & \min _{\overline{\mathbf{X}}}\quad\rho\left({\overline{\mathbf{X}}},\widehat{\mathbf{F}},\widehat{\mathbf{Q}}\right)\label{1.11a_x}\\
		\text{s.t.}  \quad 
            &\eqref{1.11d},\eqref{1.11e},\eqref{X}. \label{1.11g_x}
	\end{alignat}
\end{subequations}

\begin{subequations}
	\label{eq_Task offloading1.11_Y}
	\begin{alignat}{2}
		\mathbf{P_{1.1^{\prime}}^{\overline{Y}}}: \quad & \min _{\overline{\mathbf{Y}}}\quad\rho\left({\overline{\mathbf{Y}}},\widehat{\mathbf{F}},\widehat{\mathbf{Q}}\right)\label{1.11a_y}\\
		\text{s.t.}  \quad 
            &\eqref{1.11c},\eqref{1.11f},\eqref{Y}. \label{1.11g_y}
	\end{alignat}
\end{subequations}

\noindent Then, the optimal solution of $\mathbf{X}^r$ can be obtained by applying CVX. However, $\mathbf{X}^r$ is
a continuous variable within the closed interval of $[0,1]$ while the decision of task offloading is a binary variable. Hence, the threshold rounding method \cite{Elbassioni2021} is applied to transform the relaxed $\mathbf{X}^r$ into binary variables. Specifically, each element $x^* \in \mathbf{X}^r$ is transformed as:
\begin{equation}
       x^*=\left\{\begin{array}{l} 1, \text { if } x^* \geq \delta, \\
       0, \text { otherwise},
       \end{array}\right.
\end{equation}
\noindent where $\delta \in(0,1)$ is a positive rounding threshold. 

\par Therefore, the optimal decisions of task offloading for users can be obtained as $\mathbf{X}^*$, and problem $\mathbf{P_{1.1^{\prime}}^{\overline{X}}}$ is solved. Similarly, problem $\mathbf{P_{1.1^{\prime}}^{\overline{Y}}}$ can be solved using the analogous method. As a result, the optimal solutions of task offloading can be obtained as $\mathbf{X}^*$ and $\mathbf{Y}^*$, and thus $\mathbf{A}^r$ can be obtained.

\par However, the integrity gap is significant because the rounding process from continuous variables to binary variables may violate the constraint. Therefore, to overcome this issue after rounding, we denote $\Delta_1$ and $\Delta_2$ as the maximum violation of constraints \eqref{1.11d} and \eqref{1.11e}, respectively. As described in \cite{Feige2016}, the integrity gap $\zeta$ as follows:
\begin{equation}  
    \zeta = \min _{\overline{\mathbf{X}}} \frac{\rho}{\rho+\xi \Delta},
\end{equation}

\noindent where $\Delta=\Delta_1+\Delta_2$, and $\zeta$ is the weight of $\Delta$. Furthermore, the optimal solutions are accepted when $\zeta=1$, which is proved by Theorem \ref{the_opt_convex_gap} .
\begin{theorem}
	\label{the_opt_convex_gap}
        There is no violation of the constraints when $\zeta=1$.
\end{theorem}

\begin{myproof} Taking $\mathbf{X}^*$ as an example, constraints \eqref{1.11d} and \eqref{1.11e} are modified as follows:
\begin{subequations}
	\label{eq_gap}
	\begin{alignat}{2}
		&\sum_{u=1}^U x_a^u[n] \leq 1+\Delta_1, \\
            &x_a^u[n] T_{\mathrm{local}}[n]\leq \tau_u+\Delta_2,
	\end{alignat}
\end{subequations}
\noindent where $\Delta_1$ and $\Delta_2$ are obtained as follows:
\begin{subequations}
	\label{eq_gap1}
	\begin{alignat}{2}
        & \Delta_1=\max \left\{\sum_{u=1}^U x_a^u[n]-1, 0\right\}, \\
        & \Delta_2=\max \left\{x_a^u[n]T_{\mathrm{local}}[n]-\tau_u, 0\right\},
	\end{alignat}
\end{subequations}
\noindent where the solution of $\rho$ is obtained through relaxing the variables $\overline{\mathbf{X}}$, while the solution of $\rho+\xi \Delta$ is obtained after rounding the relaxed variables. We consider that the best rounding is achieved when $\zeta \ (\zeta \leq 1)$ is closer to 1. In other words, $\zeta=1$, when $\Delta_1=0$ and $\Delta_2=0$. Similarly, the same proof also applies to $\mathbf{Y}^*$.
\end{myproof}
}

\subsection{Computation Resource Allocation}
\label{sec_Resource Allocation}

\par Given the optimized task offloading $\mathbf{A}^r=\big\{a_{u, m}^r[n], \forall u, m, n\big\}$ for users and the UAV trajectory control $\widehat{\mathbf{Q}}=\{\widehat{\mathbf{q}}_m[n], \forall m,n\}$, the computation resource allocation problem can be expressed as:
\begin{subequations}
	\label{eq_Resource Allocatio}
	\begin{alignat}{2}
		\mathbf{P_{1.2}}: \quad &\min _{\mathbf{F}}\quad\rho\left({\mathbf{A^r}},{\mathbf{F}},\widehat{\mathbf{Q}}\right)\label{17a}\\
		\text{s.t.} \quad  
            &0 \leq f_{m,u}[n] \leq f_m^{\text{max}}, \ \forall m,n,\label{1.2b}\\
           &\sum_{u=1}^\mathbf{U_0} f_{m,u}[n] \leq f_m^{\text{max}},  \  \forall u,m,n. \label{1.2c}
	\end{alignat}
\end{subequations}

\par It can be deduced from Theorem \ref{the_opt_convex3} that  \eqref{17a} is convex. The sub-problem of computation resource allocation is solved by adopting the KKT method. This is because KKT method can effectively deal with complex problems by transforming original problems into optimization problems with equality constraints, and improve the solving efficiency when considering multiple constraints.
\begin{theorem}
	\label{the_opt_convex3}
        Problem $\mathbf{P_{1.2}}$ is a convex optimization problem.
\end{theorem}

\begin{myproof}
      The first-order derivative of $\rho$ with respect to $f_{m,u}[n]$ is calculated as:
    \begin{equation}
          \label{eq_resource_proof1}  
           \frac{\partial \rho}{\partial f_{m,u}}=-\frac{a_{u, m}^r[n]D_u[n]C_u}{f_{m,u}[n]^2}+2\kappa_m{a_{u, m}^r[n]D_u[n]C_u}f_{m,u}[n].
    \end{equation}
       \par The second-order derivative of $\rho$ with respect to $f_{m,u}[n]$ is calculated as:
    \begin{equation}
          \label{eq_resource_proof2}  
           \frac{\partial ^2\rho}{\partial f_{m,u}^2}=-\frac{2{a_{u, m}^r[n]D_u[n]C_u}}{f_{m,u}[n]^3}+2\kappa_m{a_{u, m}^r[n]D_u[n]C_u}.
    \end{equation}
      \par It is clear that the second derivative $\frac{\partial ^2\rho}{\partial f_{m,u}^2} \geq 0$, and objective function $\rho$ is a convex function with regard to $f_{m,u}[n]$. Therefore, problem $\mathbf{P_{1.2}}$ is convex \textcolor{b2}{and has an optimal solution}.
\end{myproof}

\par Hence, the slater condition is satisfied and the problem can be solved by using partial Lagrange function, which is formulated as:
\begin{equation}
\label{eq_Lagrange}
     L(F, \lambda)= \rho({\mathbf{A^r}},{\mathbf{F}},\widehat{\mathbf{Q}})+\lambda\left(\sum_{u=1}^\mathbf{U_0} f_{m,u}[n]- f_m^{\text{max}}\right),
\end{equation}

\noindent where $\lambda\geq 0$ is the Lagrange multiplier related to the computation resource constraint of UAV.

\par Furthermore, the KKT conditions are used to obtain the optimal computation resource allocation $\mathbf{F^r}$. Specifically, by differentiating $L(F, \lambda)$ with respect to $f_{m,u}^r[n]$ and setting the result as 0, which can be expressed as:
\begin{align}
      \label{eq_optimal}  
       &2\kappa_m{a_{u, m}^r[n]D_u[n]C_u}f_{m,u}^r[n]^3+\lambda^{*}f_{m,u}^r[n]^2\nonumber\\
       &-{a_{u, m}^r[n]D_u[n]C_u}=0.
\end{align}

\par Then, the optimal computation resource allocation solution $\mathbf{F^r}$ can be achieved by applying the bisection algorithm \cite{Zhao2019}, as shown in Algorithm \ref{Algorithm 1}. Specifically, the search accuracy threshold $\varepsilon$, lower bound $\lambda^{\text{min}}$ and upper bound $\lambda^{\text{max}}$ are set firstly (Line 1). Then, enter the loop of iterations until condition $\lambda^{\text{max}}-\lambda^{\text{min}}\geq \varepsilon$ is satisfied (Line 2). Furthermore, in each iteration, the algorithm determines the value of $\lambda$ by bisection and performs the calculation of the computation resource allocation $f_{m,u}^{r}[n]$, then compares it with the maximum resource owned by the UAV $f_m^{\text{max}}$ and updates $\lambda$ if the condition is met (Lines 3 to 12). 

\begin{algorithm}[]	
    \label{Algorithm 1}
    \SetAlgoLined
    \KwIn{The maximum resource owned by the UAV $f_m^{\text{max}}$.}
    \KwOut{The optimal computation resource allocation $\mathbf{F}^{r}=\{f_{m,u}^{r}[n], m\in \mathcal{M}\}$.}
    \textbf{ Initialization:} 
	Search accuracy threshold: $\varepsilon$, the lower bound $\lambda^{\text{min}}=0$ and the upper bound $\lambda^{\text{max}}=\lambda^{\text{bound}}$\;
    \While {$\lambda^{\text{max}}-\lambda^{\text{min}}\geq \varepsilon$}
    {
        Define $\lambda=\frac{\lambda^{\text{min}}+\lambda^{\text{max}}}{2}$\;
        \For {$u\in \mathbf{U}_0$}
        {
            Compute $f_{m,u}^{r}[n]$ by substituting $\lambda$ into Eq. (\ref{eq_optimal});
        }
        \eIf {$f_{m,u}^{r}[n]\geq f_m^{\text{max}}$}
        {
            $\lambda^{\text{min}}=\lambda$\;
        }
        {
            $\lambda^{\text{max}}=\lambda$\;
        }
    }
    \Return{$\mathbf{F}^{r}=\{f_{m,u}^{r}[n], m\in \mathcal{M}\}$.}
    \caption{Bisection Algorithm-based Computation Resource Allocation}
\end{algorithm}	

\subsection{UAV Trajectory Control}
\label{UAV Trajectory Control}

\par Given the optimized decisions of task offloading $\mathbf{A}^r=\left\{a_{u, m}^r[n], \forall u, m, n\right\}$ and computation resource allocation $\mathbf{F}^r=\left\{f_{m,u}^r[n], \forall u \in \mathbf{U}_0, m, n\right\}$, the UAV trajectory control problem can be expressed as:
\begin{subequations}
	\label{eq_UAV trajectory}
	\begin{alignat}{2}
		\mathbf{P_{1.3}}: \quad &\min _{\mathbf{Q}}\quad\rho\left({\mathbf{A^r}},\mathbf{F^r},{\mathbf{Q}}\right)\label{50a}\\
		\text{s.t.} \quad &\mathbf{q}_m[n+1]=\mathbf{q}_m[n]+v_m[n] \delta_t+\frac{1}{2} a_m[n] \delta_t^2,  \ \forall m, n, \label {1.3a}\\
         &v_m[n+1]=v_m[n]+a_m[n] \delta_t, \ \forall m, n,\label{1.3b}\\
           &\left\|v_m[n]\right\| \leq V_{\max }, \quad\left\|v_m[n]\right\| \geq V_{\min }, \  \forall m, n,\label {1.3c}\\
           &\left\|\mathbf{q}_m[n]-\mathbf{q}_i[n]\right\| \geq D_{\min }, \ \forall m \neq i, n,\label{1.3d}\\
            &\mathbf{q}_m[1]=\mathbf{q}_m[N], \ \forall m,\label{1.3e}\\
           &\left\|\mathbf{q}_m[n+1]-\mathbf{q}_m[n]\right\| \leq D_{\max }=V_{\max}\delta_t,  \ \forall m, n.\label{1.3f}
	\end{alignat}
\end{subequations}

\par Since the constraints \eqref{1.3c} and \eqref{1.3d} are non-convex, the sub-problem $\mathbf{P_{1.3}}$ is non-convex. In view of the fact that the SCA method shows reliable convergence when dealing with non-convex optimization problems, and its property of effectively handling complex constraints, we choose to use SCA method to solve the sub-problem of UAV trajectory control, so as to improve the efficiency and speed of problem solving. 

\par According to Lemma \ref{lemma_given}, we can transform the non-convex constraint into a lower bound on a convex function.

\begin{lemma}
    \label{lemma_given}
        For a given local point $x^r$, we have the following inequality:
    \begin{equation}
    \label{eq_formula1}  
         \textcolor{b2}{x^2 \geq 2 x^r\left(x-x^r\right)+\left(x^r\right)^2}.
    \end{equation}
\end{lemma}

\begin{myproof}
      we first define a quadratic function as follows: $f(x)=x^2$. It can be easily observed that $f(x)$ is a convex function. Since any convex function can obtain its lower bound by adopting its first-order Taylor expansion at a local point, the inequality holds in \eqref{eq_formula1} \cite{Ji2020}.
\end{myproof}

\par For the non-convex constraint \eqref{1.3c}, we first rewrite the non-convex constraints in \eqref{1.3c} as:
\begin{equation}
\label{eq_formula2}  
       \left\|v_m[n]\right\|^2 \geq V_{\min }^2, \ \left\|v_m[n]\right\|^2 \leq V_{\max }^2.
\end{equation}

\noindent Furthermore, \textcolor{b2}{according to Lemma \ref{lemma_given}}, given the local point $v_m^r[n]$, we have:
\begin{align}
\label{eq_formula3}
     \left\|v_m[n]\right\|^2 
     & \geq\left\|v_m^r[n]\right\|^2+2 {\left(v_m^r[n]\right)^T}\left(v_m[n]-v_m^r[n]\right)\nonumber\\ 
     & \triangleq \vartheta_m^r[n].
\end{align}

\noindent Therefore, the non-convex constraint \eqref{1.3c} can be bounded as:
\begin{equation}
\label{eq_formula4}
     \vartheta_m^r[n]\geq V_{\min }^2,\ \vartheta_m^r[n] \leq V_{\max }^2.
\end{equation}


\par For the non-convex constraint \eqref{1.3d}, we introduce an auxiliary variable $D_{m,i}[n]$ as:
\begin{equation}
\label{eq_ayxiliary_dmi}
     D_{m,i}[n] = \left\|\mathbf{q}_m[n] - \mathbf{q}_i[n]\right\|, \ \forall n, m \neq i.
\end{equation}

\noindent Furthermore, \textcolor{b2}{since the distance between any two UAVs in the system should be greater than or equal to the defined safe distance,} Eq.~\eqref{eq_ayxiliary_dmi} can be further rewritten as $\big\|D_{m, i}\big\|^2 \geq D_{\min }^2$. Thus, given the fixed point $D^r_{m,i} [n]$, we have:
\label{eq_formula5}
      \begin{align}
      \left\|D_{m, i}[n]\right\|^2 & \geq 2\left(D_{m, i}^r[n]\right)^T\left(D_{m, i}[n]\right.\left.-D_{m, i}^r[n]\right)\nonumber\\ 
      & +\left\|D_{m, i}^r[n]\right\|^2.
\end{align}

\noindent Therefore, for the given points of UAV $\mathbf{q}^r_i [n]$ and $\mathbf{q}^r_m[n]$, the non-convex \textcolor{b2}{constraint} \eqref{1.3d} can be converted to:
\begin{align}
\label{eq_formula6}
      D_{\min }^2 & \leq \left\|\mathbf{q}_m^r[n]+\mathbf{q}_i^r[n]\right\|^2+2\left(\mathbf{q}_m^r[n]-\mathbf{q}_i^r[n]\right)^T \nonumber\\
      &\times\left(\mathbf{q}_m[n]-\mathbf{q}_i[n]-\mathbf{q}_m^r[n]+\mathbf{q}_i^r[n]\right)\nonumber\\ 
      & \triangleq \zeta_{m, i}^r[n], \quad \forall n,m \neq i.
\end{align}

\par In conclusion, according to \eqref{eq_formula4} and \eqref{eq_formula6}, sub-problem $\mathbf{P_{1.3}}$ can be transformed into a convex problem $\mathbf{P_{1.3^{\prime}}}$ by introducing the auxiliary variables $\vartheta_m^r[n]$ and $\zeta_{m, i}^r[n]$, which is as follows:
\begin{subequations}
	\label{eq_problem 1.3'}
	\begin{alignat}{2}
		\mathbf{P_{1.3^{\prime}}}: \quad &\min _{\mathbf{Q}}\quad\rho\left({\mathbf{A^r}},{\mathbf{F^r}},{\mathbf{Q}}\right)\label{26a}\\
		\text{s.t.} \quad & \vartheta_m^r[n] \geq V_{\min}^2, \ \vartheta_m^r[n] \leq V_{\max }^2, \ \forall m, n,\label {26b}\\
             &\zeta_{m, i}^r[n] \geq D_{\min }^2, \ \forall n, m \neq i,\label {26c}\\
	    &\eqref{1.3a},\eqref{1.3b},\eqref{1.3e},\eqref{1.3f}.\label{26d}
	\end{alignat}
\end{subequations}

\noindent It can be directly solved by existing optimization software such as CVX \cite{Grant2014}.
{\color{b2}
\begin{theorem}
	\label {the_opt_convex2}
 Problem $\mathbf{P_{1.3^{\prime}}}$ does not change the optimality of problem $\mathbf{P_{1.3}}$, i.e., $\rho_{1.3^{\prime}}\left({\mathbf{A^r}},{\mathbf{F^r}},{\mathbf{Q}}\right) \leq \rho_{1.3}\left({\mathbf{A^r}},{\mathbf{F^r}},{\mathbf{Q}}\right)\nonumber$.
\end{theorem}

\begin{myproof}
     By applying the lower bound results in \eqref{eq_formula4} and \eqref{eq_formula6}, it can be deduced that the set of feasible solutions for subproblem $\mathbf{P_{1.3^{\prime}}}$ is included within that for $\mathbf{P_{1.3}}$. Accordingly, the objective function value obtained from $\mathbf{P_{1.3^{\prime}}}$ is always less than or equal to that from $\mathbf{P_{1.3}}$, which indicates that there exists at least one local optimal solution\cite{Ji2020}.
\end{myproof}
}
\subsection {Main Steps of JTORATC and Analysis}
\par In this section, the main steps of the JTORATC are shown in Algorithm \ref{Algorithm 2}. First, the decision parameters of task offloading, computation resource allocation, UAV trajectory, and iteration numbers are initialized (Line 1). Then, the optimal decisions of task offloading, computation resource allocation and UAV trajectory control are obtained by solving the sub-problems $\mathbf{P_{1.1}}$, $\mathbf{P_{1.2}}$ and $\mathbf{P_{1.3}}$ alternately (Lines 2 to 8). In addition, the \textcolor{b2}{convergence} and computational complexity of JTORATC are proved respectively. 

\begin{algorithm}	
    \label{Algorithm 2}
    \SetAlgoLined
    \KwIn{Task offloading $\mathbf{A}^0$, computation resource allocation $\mathbf{F}^0$, UAV trajectory $\mathbf{Q}^0$, and iteration index $r^0$.}
    \KwOut{Optimal task offloading $\mathbf{A}^{opt}$, computation resource allocation $\mathbf{F}^{opt}$ and UAV trajectory $\mathbf{Q}^{opt}$.}
    \textbf{Initialization:} 
    The iteration index $r=1$, the tolerance of accuracy $\epsilon=10^{-4}$, $\mathbf{A}^0=\emptyset$;\\
    \Repeat{$\frac{\rho_{r+1}-\rho_r}{\rho_r} \leq \epsilon$}
    {
         Fix $\left\{{\mathbf{F}^r, \mathbf{Q}^r}\right\}$, obtain the optimal solution of the problem $\mathbf{P_{1.1}}$ is ${\mathbf{A}^{r+1}}$\;
         Fix $\left\{{\mathbf{A}^{r+1}},\mathbf{F}^r,\mathbf{Q}^r\right\}$, obtain the optimal solution of the problem $\mathbf{P_{1.2}}$ based on Algorithm 1 is ${\mathbf{F}^{r+1}}$\;
         Fix $\left\{{\mathbf{A}^{r+1}},\mathbf{F}^{r+1},\mathbf{Q}^r\right\}$, obtain the optimal solution of the problem $\mathbf{P_{1.3}}$ is ${\mathbf{Q}^{r+1}}$\;
         
         Update $\left\{{{\mathbf{A},\mathbf{F},\mathbf{Q}}}\right\}^r\leftarrow\ \left\{{\mathbf{A}^{r+1}, \mathbf{F}^{r+1},\mathbf{Q}^{r+1}}\right\}$ and $r=r+1$\;
    }
    \Return{$\left\{ \mathbf{A}^{opt}, \mathbf{F}^{opt}, \mathbf{Q}^{opt} \right\}$.}
    \caption{JTORATC}
\end{algorithm}

{\color{b2}
\subsubsection {Convergence and Computation Complexity}
\label{sec_Convergence Complexity}

\par The convergence of JTORATC is analyzed as follows:
\begin{subequations}
   \begin{alignat}{2}
\rho\left(\mathbf{A}^{r}, \mathbf{F}^{r}, \mathbf{Q}^r\right) & \stackrel{(a)}{\geq} \rho\left(\mathbf{A}^{r+1}, \mathbf{F}^{r}, \mathbf{Q}^r\right), \label{a}\\
& \stackrel{(b)}{\geq} \rho\left(\mathbf{A}^{r+1}, \mathbf{F}^{r+1}, \mathbf{Q}^{r}\right), \label{b}\\
& \stackrel{(c)}{\geq} \rho\left(\mathbf{A}^{r+1}, \mathbf{F}^{r+1}, \mathbf{Q}^{r+1}\right).\label{c} 	\end{alignat}
\end{subequations}

\noindent Since task offloading is solved for a given resource allocation and UAV trajectory, the inequality \eqref{a} holds. Furthermore, inequality \eqref{b} is satisfied because the KKT condition is introduced to solve resource allocation sub-problem. Moreover, the inequality \eqref{c} is due to the sub-optimality of UAV trajectory. Besides, the objective function $\rho\left(\mathbf{A}, \mathbf{F}, \mathbf{Q}\right)$ is always positive owing to its non-negative expression. Therefore, the objective function is always non-increasing after every iteration, which is also ﬁnitely lower-bounded. 

\par The computational complexity of problem $\mathbf{P_{1.1}}$ is $\mathcal{O}\left(K\left(\log_2(\frac{1}{\epsilon}+1\right)\right)$ \cite{Hong2020} based on the analysis of Theorem \ref{the_opt_convex1} and Theorem \ref{the_opt_convex_gap}, where $K=U+M$, $U$ is the number of users, $M$ is the number of UAVs, and $\epsilon$ is the tolerance of accuracy. Furthermore, problem $\mathbf{P_{1.2}}$ is solved based on the bisection algorithm presented in Algorithm 1, and its computational complexity can be calculated as $\mathcal{O}\left(\log_2\left(\frac{\lambda^{\max }-\lambda^{\min }}{\varepsilon}\right)\right)$ \cite{Zhao2019}, where $\lambda^{\min }$ and $\lambda^{\max }$ are the lower and upper bounds of $\lambda$, respectively, and $\varepsilon$ denotes the search accuracy. Moreover, problem $\mathbf{P_{1.3}}$ involves solving non-convex problem, and the computational complexity is $\mathcal{O}(M^{3.5}\log_2(\frac{1}{\epsilon}))$ \cite{Wang2022}. Therefore, the proposed JTORATC approach converges in polynomial, and the computational complexity is $\mathcal{O}\big(I_c\big(K\big(\log_2(\frac{1}{\epsilon}+1\big)+\log_2\big(\frac{\lambda^{\max }-\lambda^{\min }}{\varepsilon}\big)+M^{3.5}\log_2(\frac{1}{\epsilon})\big)\big)$, where $I_c$ is the number of outer iterations of Algorithm \ref{Algorithm 2}.
}


\section{Simulation Results}
\label{sec_results}
\par In this section, simulation results are presented to validate the effectiveness of the proposed approach.

\subsection{Simulation Setup}
\label{sec_Simulation_Setup}

\par  We perform simulations to verify the validity of our proposed JTORATC method. Specifically, all the simulations are performed in MATLAB R2022b on a desktop computer equipped with an Intel(R) Core(TM) i7-8750H CPU @ 2.20GHz 2.21GHz and 8 GB RAM.

\par \textbf{Scenarios.} We consider a multi-UAV-assisted MEC system where 2 UAVs, 4 UAVs and 6 UAVs  are deployed to offer offloading services to 8 users in a $2500\times 3000 \ \text{m}^2$ rectangular area. The timeline is set as $T = 100$ s which is divided into $50$ time slots.

\par \textbf{Parameters.} The default values of the simulation parameters are listed in Table \ref{tab_simuParameter}. 

\begin{table}[!hbt]
	\caption{Simulation parameters}
	\label{tab_simuParameter}
	\renewcommand*{\arraystretch}{1}
	\begin{center}
		\begin{tabular}{p{.06\textwidth}|p{.21\textwidth}|p{.13\textwidth}}
			\hline
			\hline
			\textbf{Symbol}&\textbf{Meaning}&\textbf{Default value}\\
			\hline
				$f_u$&Computing capability of user $u$& 340 MHz\\
			\hline
				$p_u$ &Transmission power of user $u$& 30 dBm\\
                \hline
                $\tau_u$ &Deadline of the task &[0.1, 75]s\\
			\hline
                $H$ &Fixed flight altitude of UAV $m$& 100 m\\
                \hline
				$\textbf{q}_{1}^{I}$& Initial position of UAV1& $[800,1200]$ \\
				\hline
                $\textbf{q}_{2}^{I}$& Initial position of UAV2& $[2000,1000]$ \\
			\hline
				$V_{min}$ &Minimum speed of UAV $m$&20 m/s\\
			\hline
				$V_{max}$ &Maximum speed of UAV $m$&60 m/s \\
			\hline
				$f_m^{\text{max}}$ &Computing capability of UAV $m$ & 1200 MHz \\
			\hline
				$\kappa_m$ &Effective switching capacitance of UAV $m$ & $10^{-27}$\\
			\hline
				$D_{min}$& Safe distance between UAVs& 10 m\\
			\hline
				$B$ & Channel bandwidth& $20$ MHz\\
			\hline
                $\sigma^2$ &Noise power&-100 dBm\\
                \hline
		\end{tabular}
	\end{center}
\end{table} 

\par \textbf{Benchmarks.} This work evaluates the proposed JTORATC in comparison with the following schemes.

\begin{itemize}
	\item  \textit{Random offloading and JTORATC-based resource allocation and trajectory control (ROJRATC)}: the task offloading strategies of users are decided randomly, while the computation resource allocation  and the trajectories of UAVs are decided based on the proposed JTORATC.
	\item  \textit{Nearby offloading and JTORATC-based resource allocation and trajectory control (NOJRATC)}: the tasks of each user are offloaded to the nearest UAV, while the computation resource allocation and the trajectory control are decided based on the proposed JTORATC.
	\item  \textit{Many-to-many matching-based offloading and JTORATC-based resource allocation and trajectory control (MOJRATC)}: the task offloading strategies are decided by using a matching-to-matching mechanism\cite{Zhao2023}, while the computation resource allocation and the trajectories of UAVs are decided based on the proposed JTORATC.
       \item \textit{Equalizing resources, JTORATC-based offloading and trajectory control (ERJOTC)}: the computation resource allocation of UAVs are decided averagely while the strategies of task offloading and trajectory control are determined based on the proposed JTORATC.
       \item \textit{JTORATC-based offloading and resource allocation, circular trajectory (JORACT)}: the task offloading strategies and the computation resource allocation are decided based on the proposed JTORATC, while the UAVs fly following the circular trajectories.
	\item \textit{JTORATC-based offloading and resource allocation, predefined trajectory (JORAPT)}: the task offloading strategies and the computation resource allocation are decided based on the proposed JTORATC, while the UAVs follow predefined trajectories.
\end{itemize}

\subsection{Evaluation Results}
\label{sec_Evaluation Results}
\par In this section, we first present simulation results to evaluate the performance of our proposed JTORATC. Then, we compare the effects of different scenario setups on the performance of the proposed JTORATC and benchmark schemes.

\subsubsection{System Performance}

\begin{figure*}[!hbt] 
    \centering
	\subfigure[]
	{
		\begin{minipage}[t]{0.23\linewidth}
			\centering
		\includegraphics[width=1.64in]{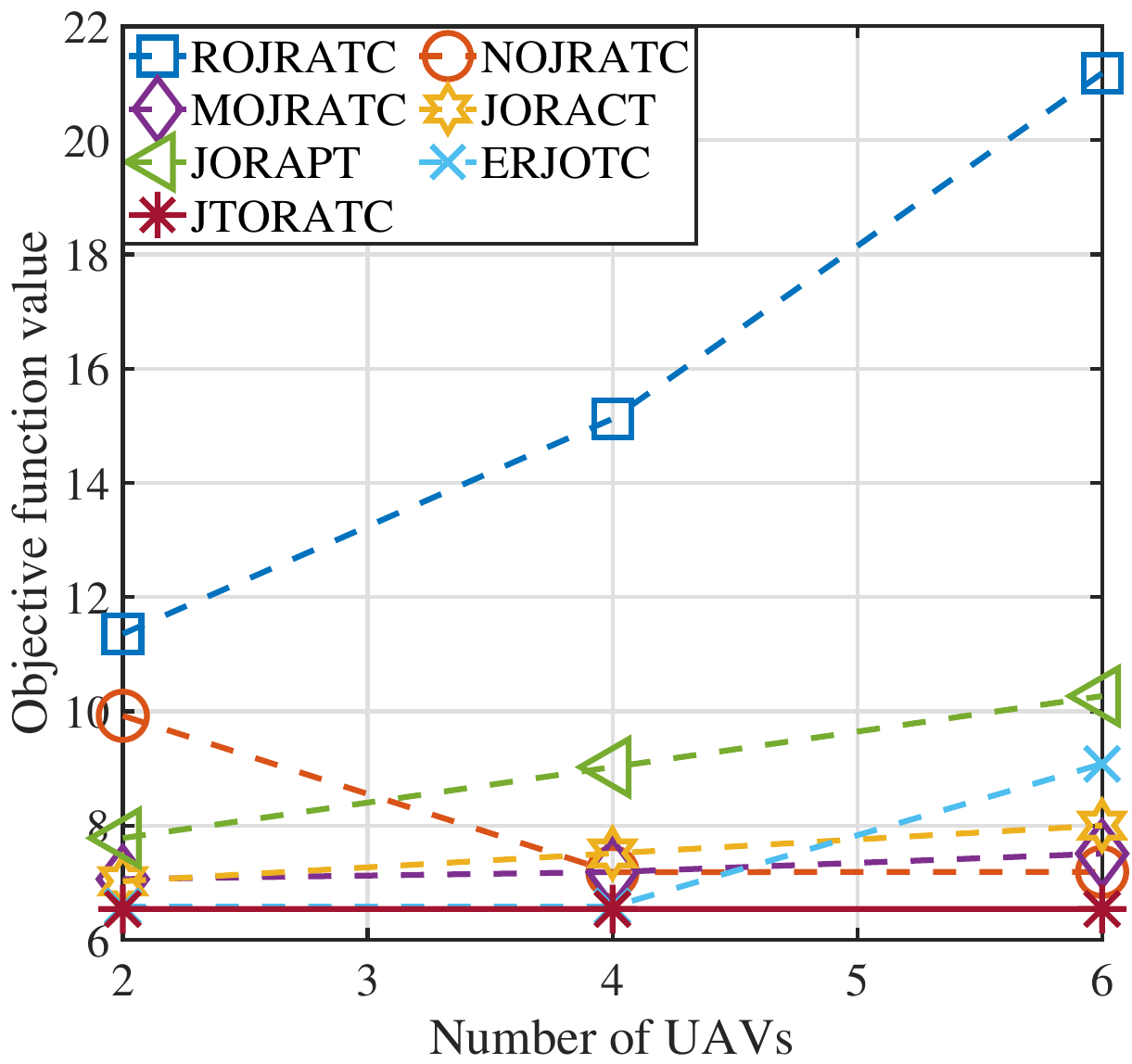}
		\end{minipage}
	}
	\subfigure[]
	{
		\begin{minipage}[t]{0.23\linewidth}
			\centering
		\includegraphics[width=1.64in]{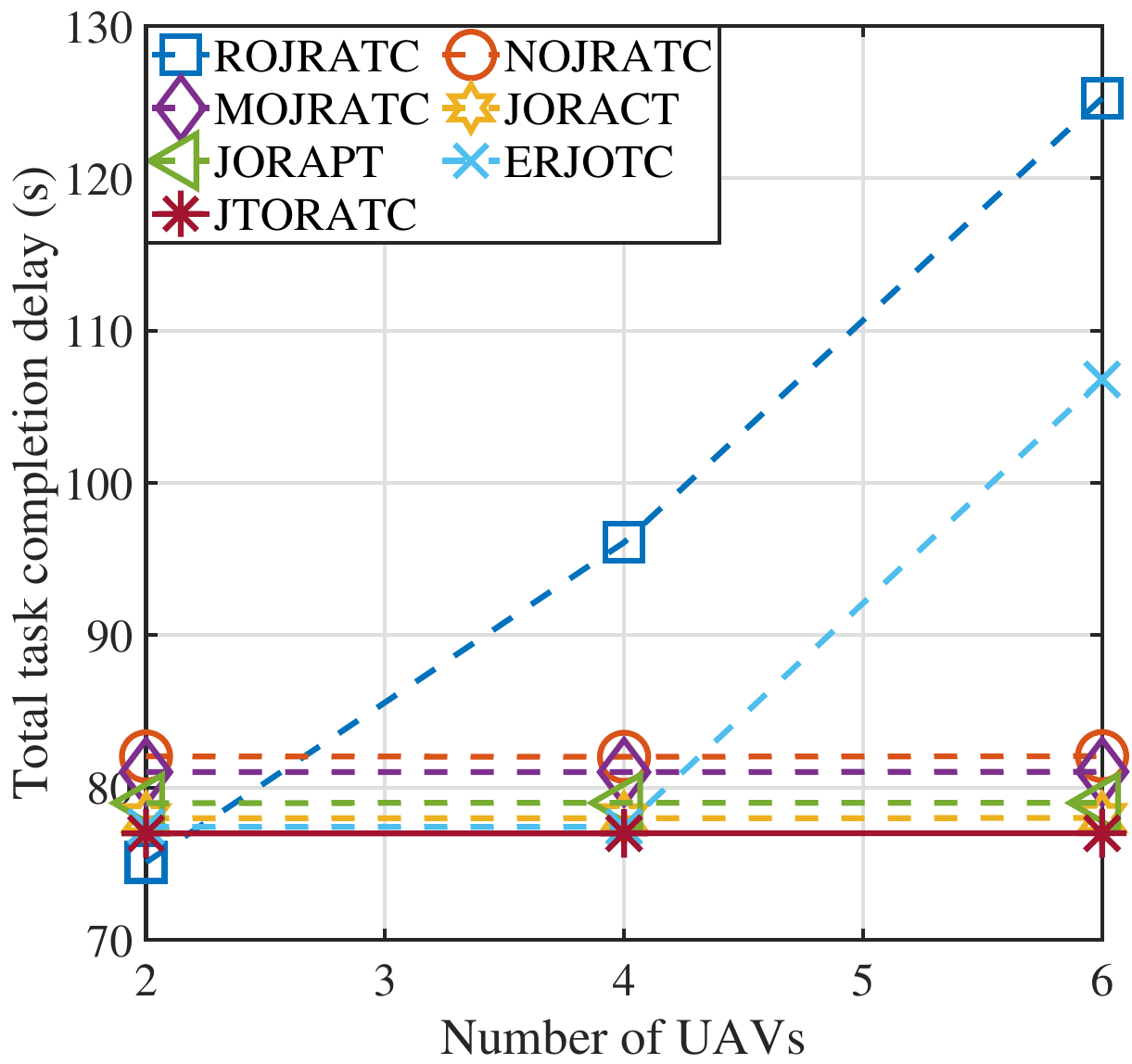}	
		\end{minipage}
	}
	\subfigure[]
	{
		\begin{minipage}[t]{0.23\linewidth}
			\centering
		\includegraphics[width=1.64in]{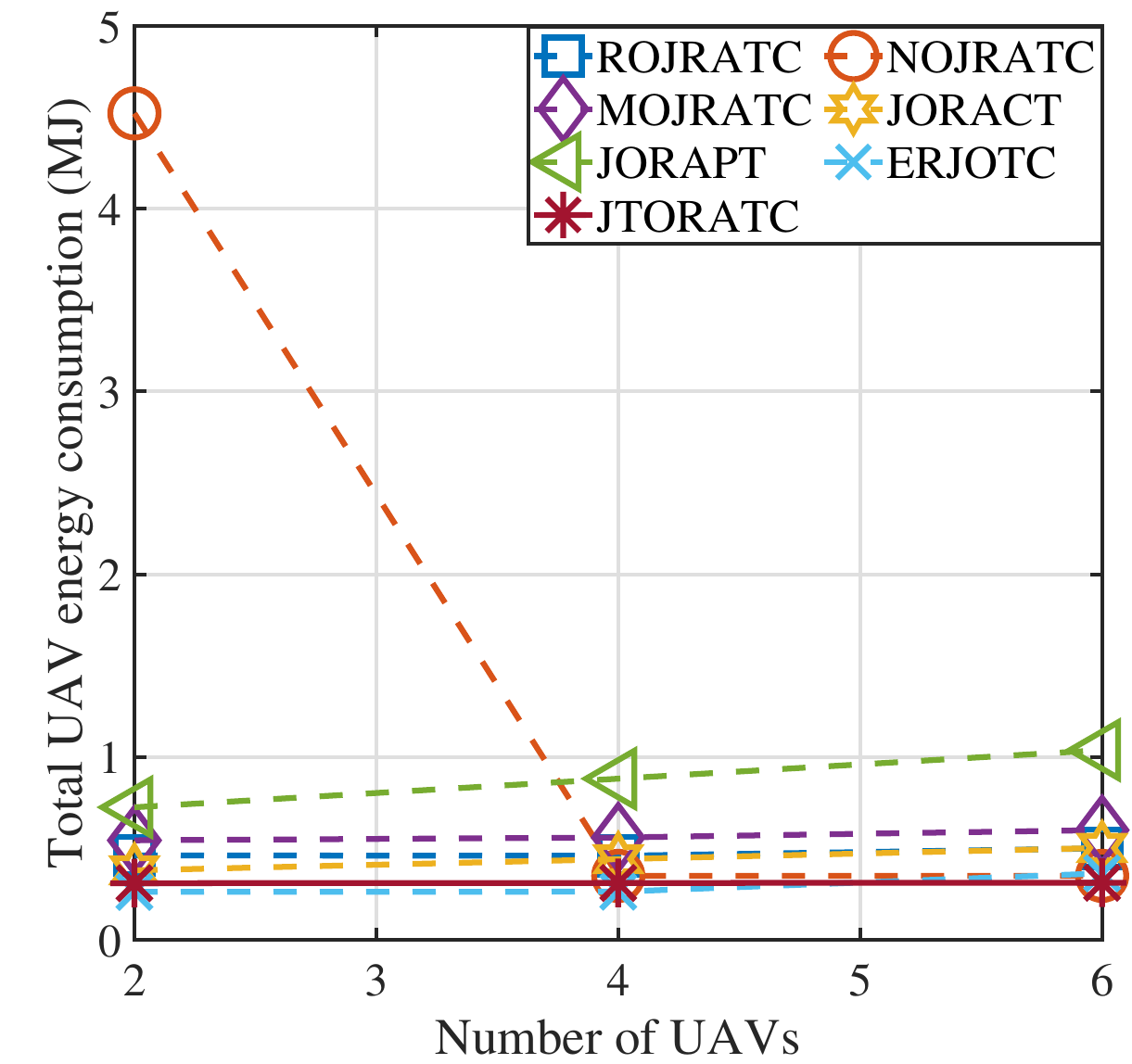}
		\end{minipage}
	}
        \subfigure[]
	{
		\begin{minipage}[t]{0.23\linewidth}
			\centering
		\includegraphics[width=1.64in]{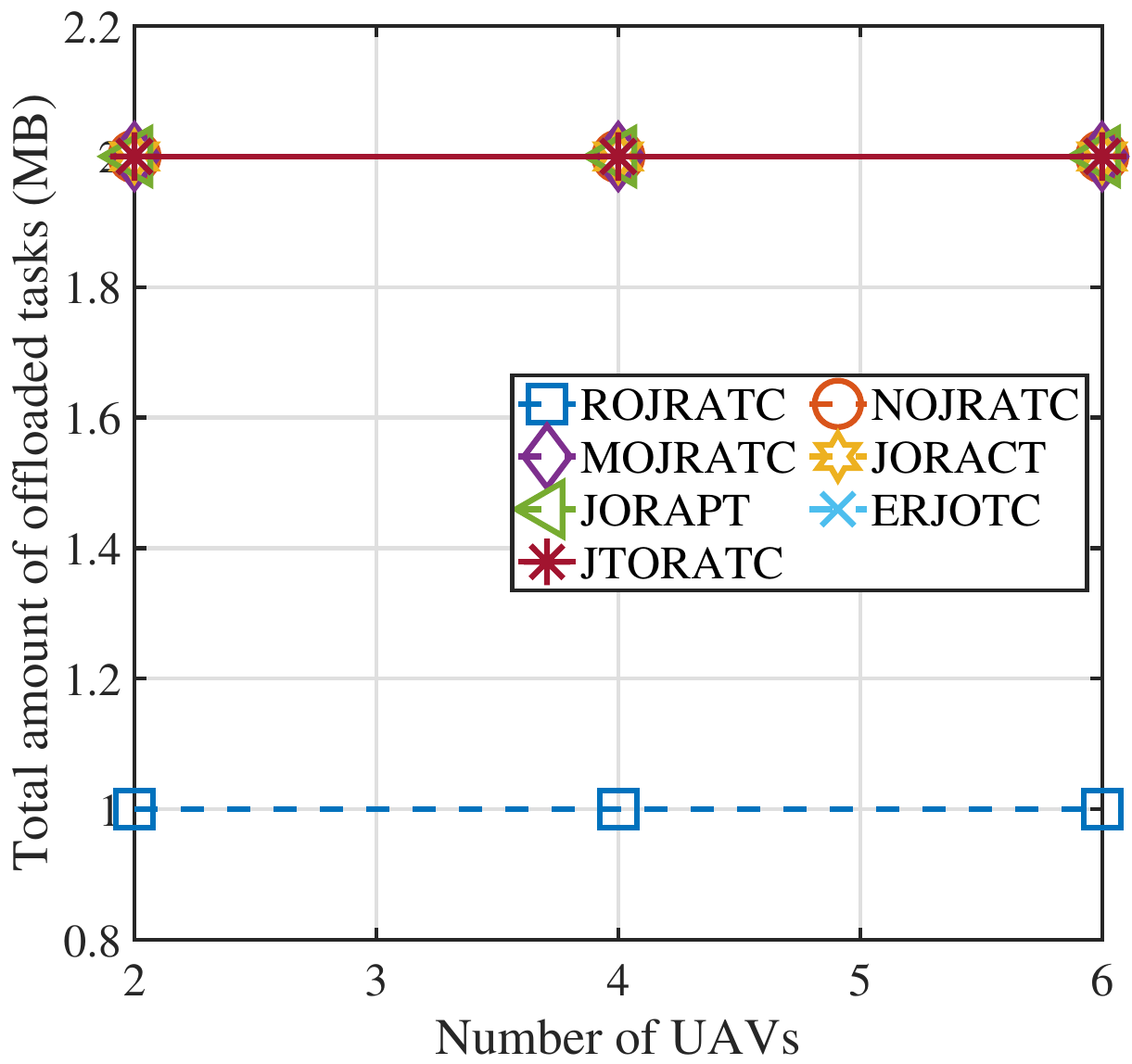}
		\end{minipage}
	}
	\centering
	\caption{System performance with UAV computation capacity. (a) Objective function value. (b) Total task completion delay. (c) Total UAV energy consumption. (d) Total amount of offloaded tasks.}
	\label{fig_System_Performance}
\end{figure*}

\par Figs. \ref{fig_System_Performance}(a), \ref{fig_System_Performance}(b), \ref{fig_System_Performance}(c), and \ref{fig_System_Performance}(d) \textcolor{b2}{show} the comparison results of the objective function value, total task completion delay, total UAV energy consumption, and total amount of offloaded tasks of seven approaches with the increasing of the number of UAVs. It can be observed from Fig. \ref{fig_System_Performance} that as the number of UAVs increases, the objective function value and total UAV energy consumption of all approaches except the NOJRATC approach increase. This is because that the increase in the number of UAVs may lead to complex task scheduling, fierce competition for computing resources and cross-conflict of trajectories. However, considering the spatio-temporal relationship of tasks, the NOJRATC simplifies task scheduling, effectively uses resources and reduces conflicts on flight trajectories.

\par Moreover, the proposed JTORATC maintains superior performance compared to other approaches as the number of UAVs increases, and this can be attributed to several reasons as follows. First, ROJRATC, NOJRATC and MOJRATC mainly focus on optimizing the task offloading. The random offloading of ROJRATC, the offloading of the nearest MEC server of NOJRATC, and the matching offloading of MOJRATC may lead to inflexibly resource allocation of the MEC servers, triggering congestion and excessive resource usage. Second, ERJOTC mainly focuses on optimizing the computation resource allocation. It can be observed from Fig. \ref{fig_System_Performance}(c) that as the number of UAVs increases, ERJOTC is slightly better than the proposed JTORATC in terms of total UAV energy consumption, but far worse than the proposed JTORATC in terms of \textcolor{b2}{the objective} function value and total task completion delay. This is because the proposed JTORATC allows for more flexible task collaborative optimization, adaptation to different task requirements and dynamic changes, as well as latency reduction through parallel processing, thus improving the overall system performance, while ERJOTC can avoid excessive concentration of resources on a few UAVs. Furthermore, JORACT and JORAPT mainly focuses on the UAV trajectory control. The circular trajectory of JORACT and predefined trajectory of JORAPT have poor performance compared with the proposed JTORATC. This is because the proposed JTORATC allows dynamic adjustment of paths to accommodate real-time changes, resulting in more efficient and flexible task execution. 
\par Accordingly, it can be concluded that the proposed JTORATC achieves optimal performance in multi-UAV-assisted MEC systems compared to other approaches.

\subsubsection{Performance Evaluation}

\par In this section, we evaluate the impact of different parameters on the system performance based on two UAVs. In this process, we keep the rest of parameters \textcolor{b2}{with} default values and only vary the UAV computation capacity, task computation intensity, task size, and number of users. Through such exploration, we further verify that our proposed JTORATC approach has significant robustness and scalability in the system.

\begin{figure*}[!hbt] 
	\centering
	\subfigure[]
	{
		\begin{minipage}[t]{0.23\linewidth}
			\centering
		\includegraphics[width=1.64in]{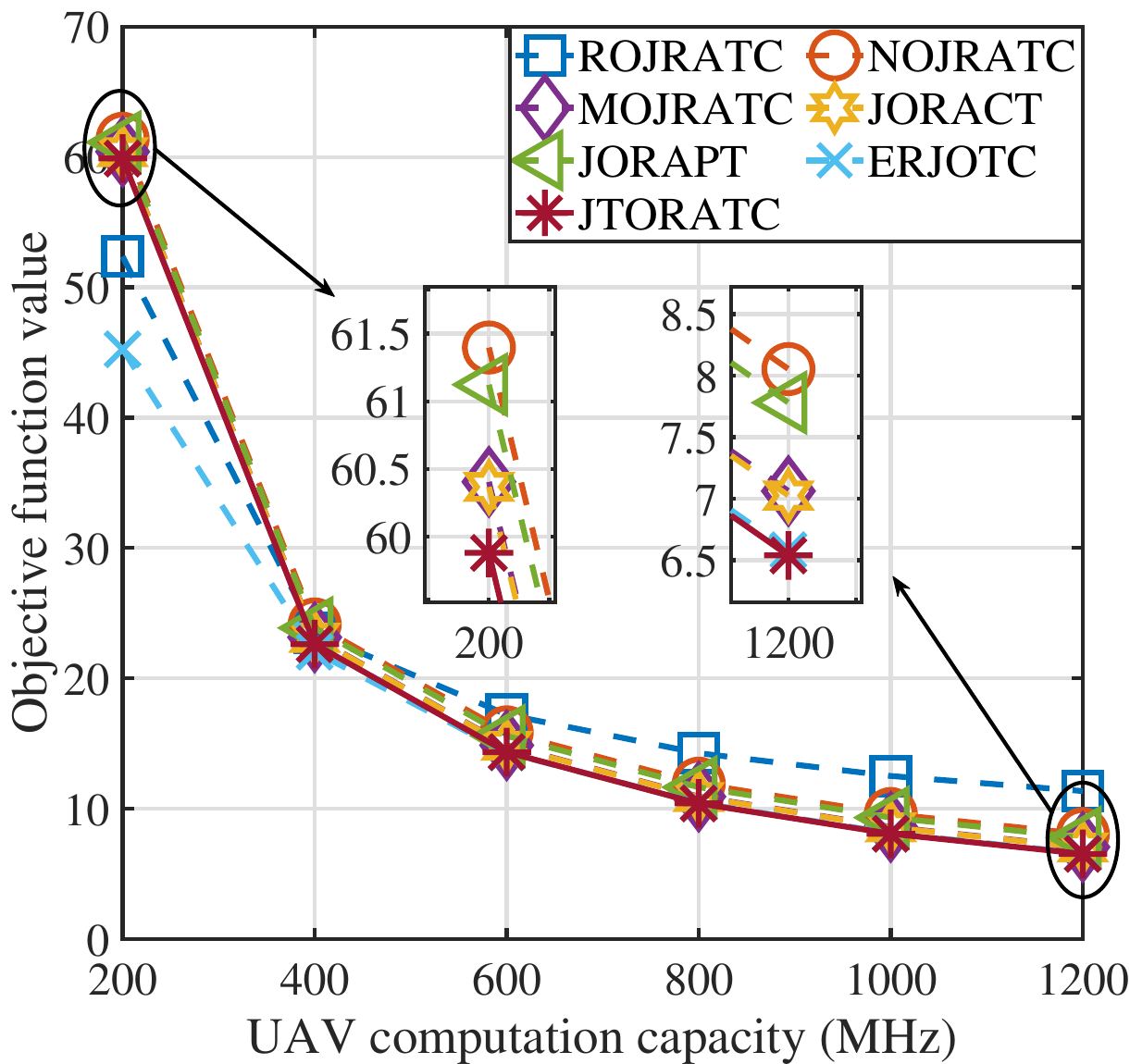}
		\end{minipage}
	}
	\subfigure[]
	{
		\begin{minipage}[t]{0.23\linewidth}
			\centering
		\includegraphics[width=1.64in]{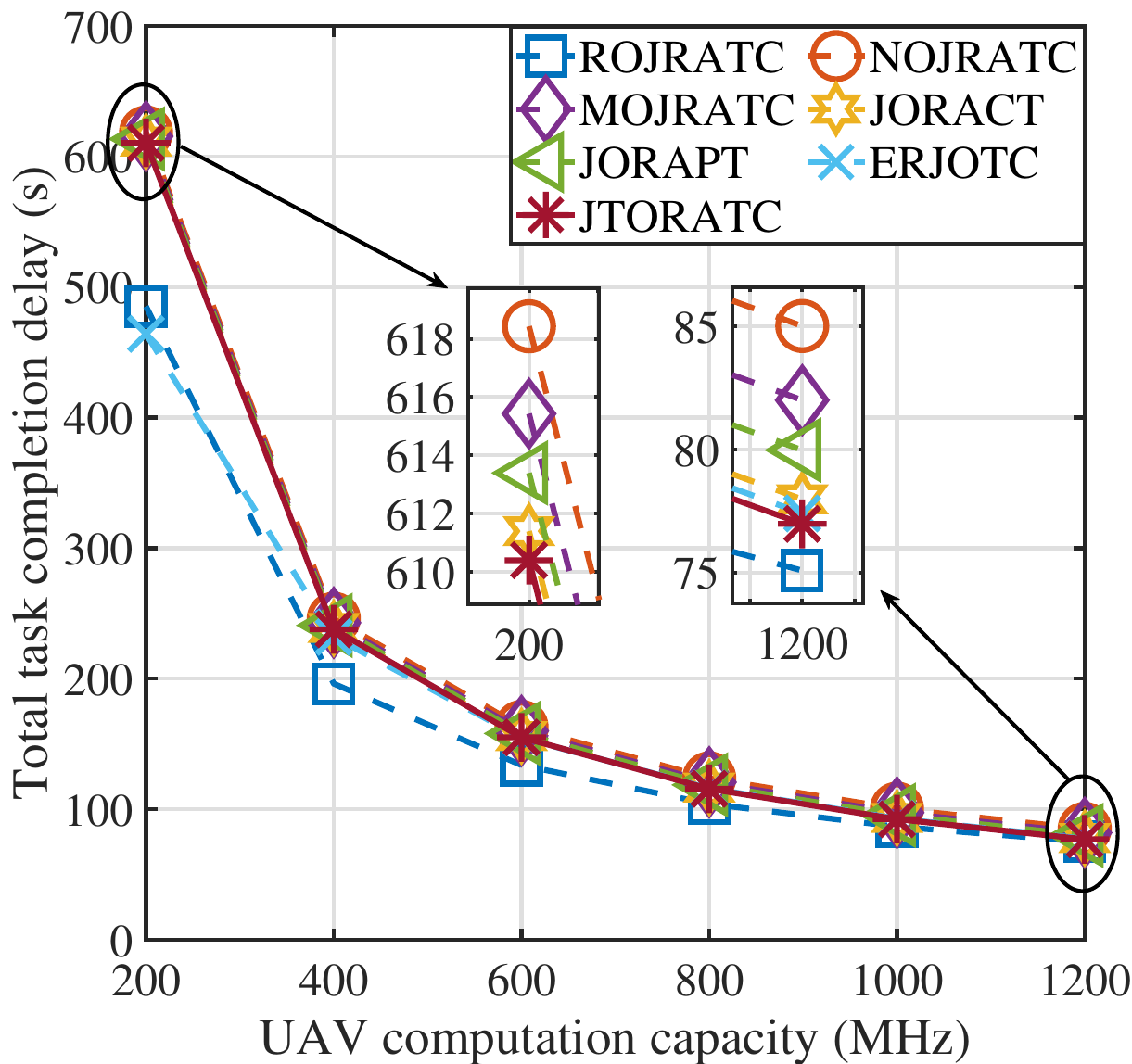}	
		\end{minipage}
	}
	\subfigure[]
	{
		\begin{minipage}[t]{0.23\linewidth}
			\centering
		\includegraphics[width=1.64in]{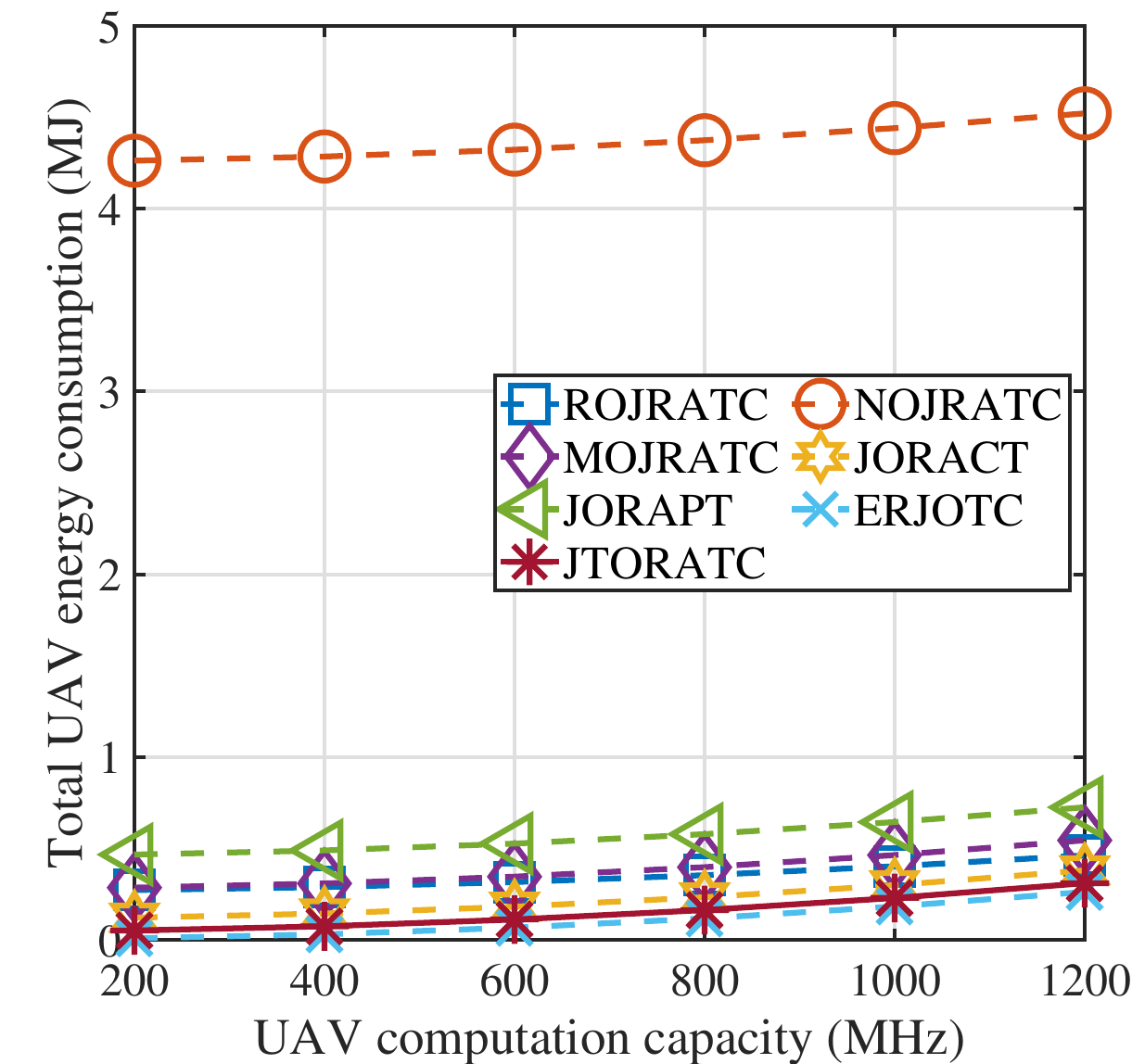}
		\end{minipage}
	}
        \subfigure[]
	{
		\begin{minipage}[t]{0.23\linewidth}
			\centering
		\includegraphics[width=1.64in]{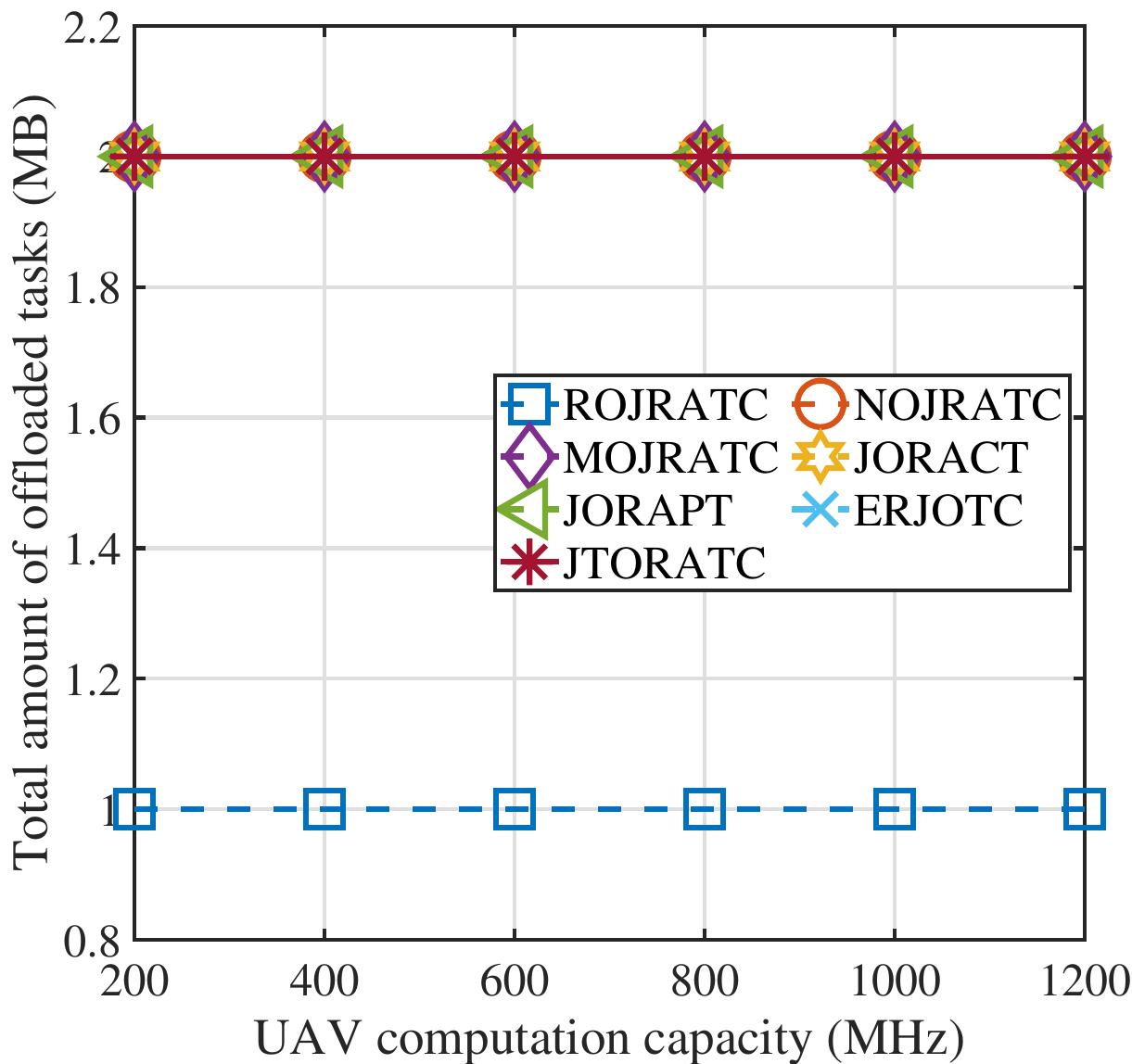}
		\end{minipage}
	}
	\centering
	\caption{System performance with numbers of UAVs. (a) Objective function value. (b) Total task completion delay. (c) Total UAV energy consumption. (d) Total amount of offloaded tasks.}
	\label{fig_computation_capacity}
\end{figure*}

\textbf{\textit{(1) Impact of UAV Computation Capacity.}} Figs. \ref{fig_computation_capacity}(a), \ref{fig_computation_capacity}(b), and \ref{fig_computation_capacity}(c) show the effects of UAV computation capacity on the objective function value, total task completion delay, total UAV energy consumption, and total amount of offloaded tasks. It can be observed from Fig. \ref{fig_computation_capacity}(a) that when the UAV computation capacity is less than 400 MHz, the ERJOTC approach achieves a lower objective function value than the proposed JTORATC. However, it is essential to recognize that while ERJOTC may find a local optimal solution in certain cases, ERJOTC cannot guarantee a global optimal solution. Therefore, this discrepancy is \textcolor{b2}{expected}. In addition, as the computation capacity of UAVs increases, the proposed JTORATC is superior to other schemes in terms of objective function. The reason is that JTORATC jointly optimizes task offloading, computing resource allocation, and UAV trajectory control to adapt the computing capabilities of different UAV, resulting in satisfactory performance.

\par From Figs. \ref{fig_computation_capacity}(b) and \ref{fig_computation_capacity}(c), it is worth noting that when the UAV computation capacity is less than 600 MHz, the ROJRATC approach obtains a lower total task completion delay than the proposed JTORATC. Similarly, this outcome is expected since the proposed JTORATC is constrained by insufficient computing resources, leading to longer task completion times. Besides, with \textcolor{b2}{an increase in the} UAV computation capacity, ROJRATC achieves a similar total completion delay to the proposed JTORATC by using the random offloading approach. However, the excessive focus on load balancing in ROJRATC causes higher energy consumption, leading to the relatively elevated energy consumption of UAVs.

\par From Fig. \ref{fig_computation_capacity}(d), we can see that for the total amount of offloaded tasks, the seven approaches all show a constant trend regardless of the varying UAV computation capacity, since the total amount of tasks completed by the UAV depends on the amount of tasks generated by the users, and has nothing to do with the computation capacity of UAV.

\par In summary, the results in Fig. \ref{fig_computation_capacity} illustrate the adaptability of the proposed JTORATC with varying computing capacities of UAVs.

\begin{figure*}[!hbt] 
	\centering
	\subfigure[]
	{
		\begin{minipage}[t]{0.23\linewidth}
			\centering
		\includegraphics[width=1.64in]{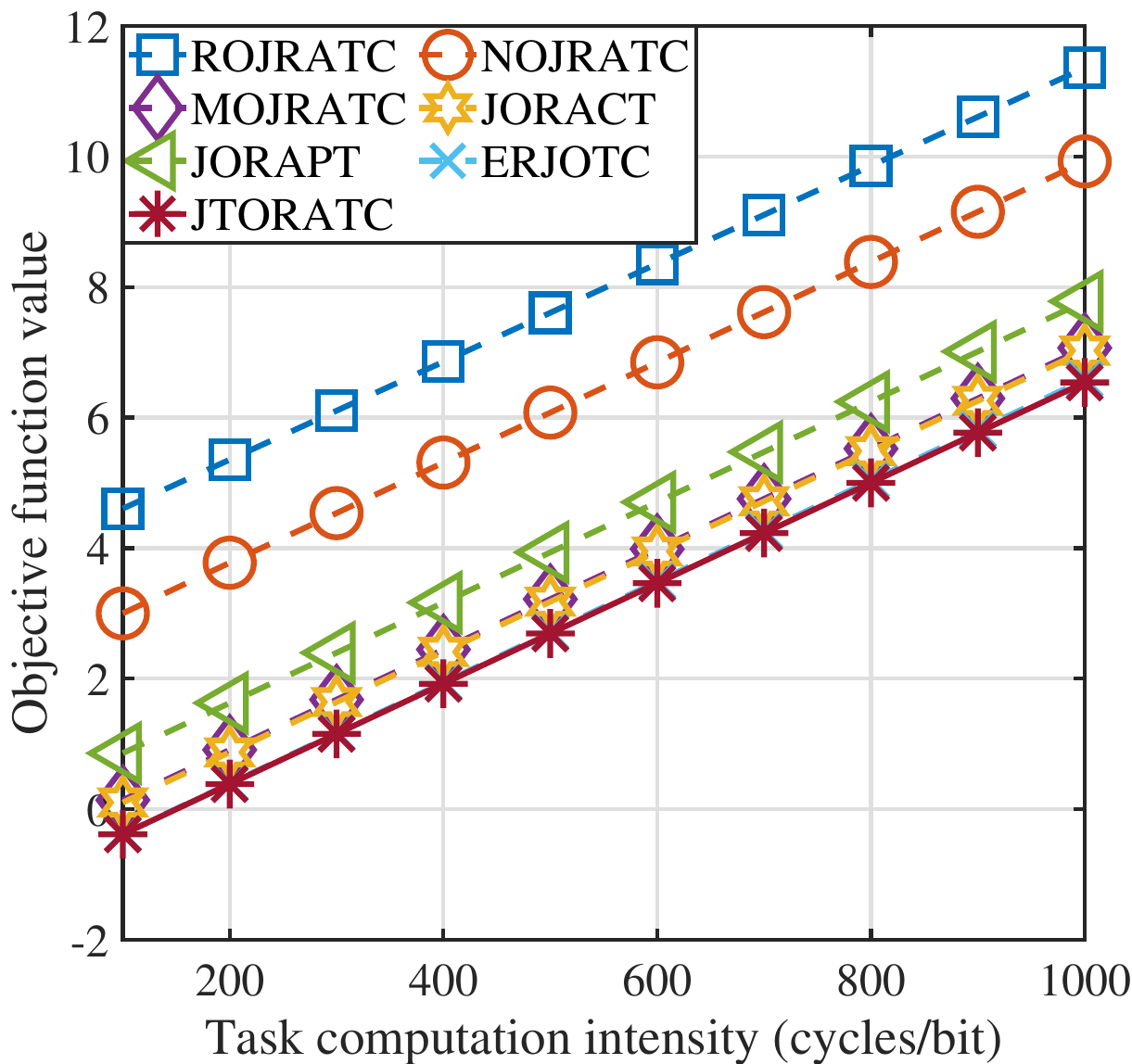}
		\end{minipage}
	}
	\subfigure[]
	{
		\begin{minipage}[t]{0.23\linewidth}
			\centering
		\includegraphics[width=1.64in]{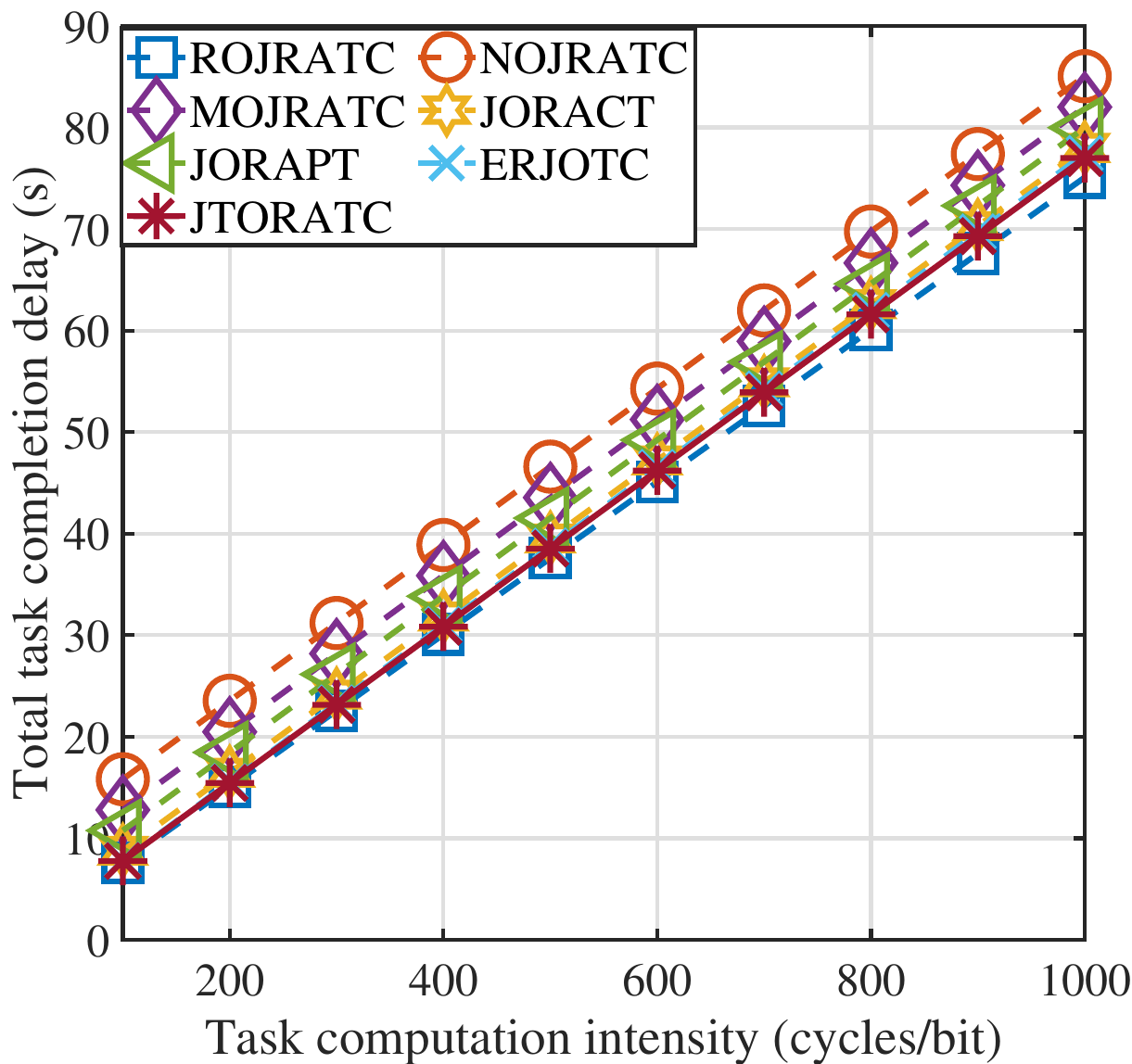}	
		\end{minipage}
	}
	\subfigure[]
	{
		\begin{minipage}[t]{0.23\linewidth}
			\centering
		\includegraphics[width=1.64in]{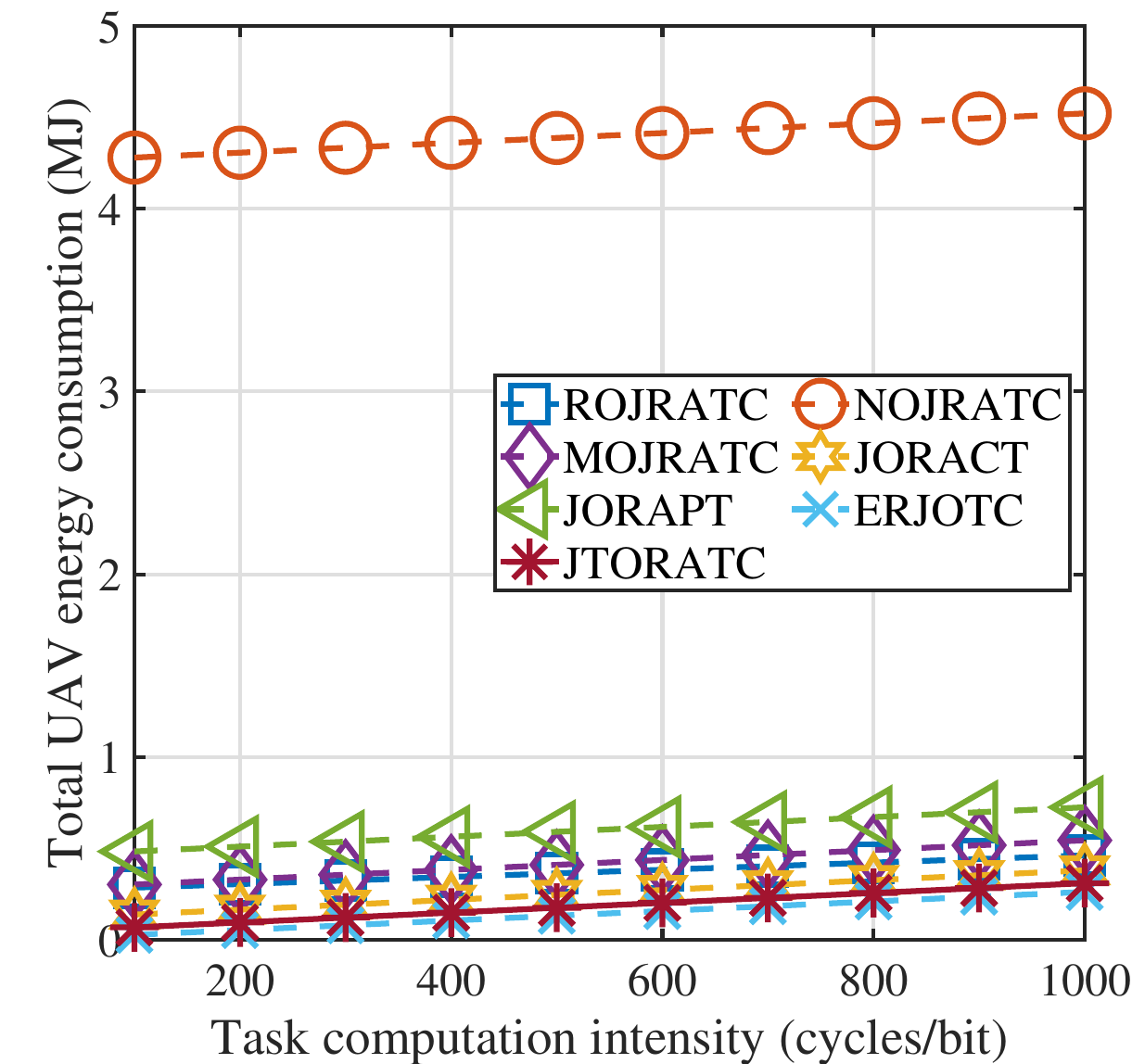}
		\end{minipage}
	}
        \subfigure[]
	{
		\begin{minipage}[t]{0.23\linewidth}
			\centering
		\includegraphics[width=1.64in]{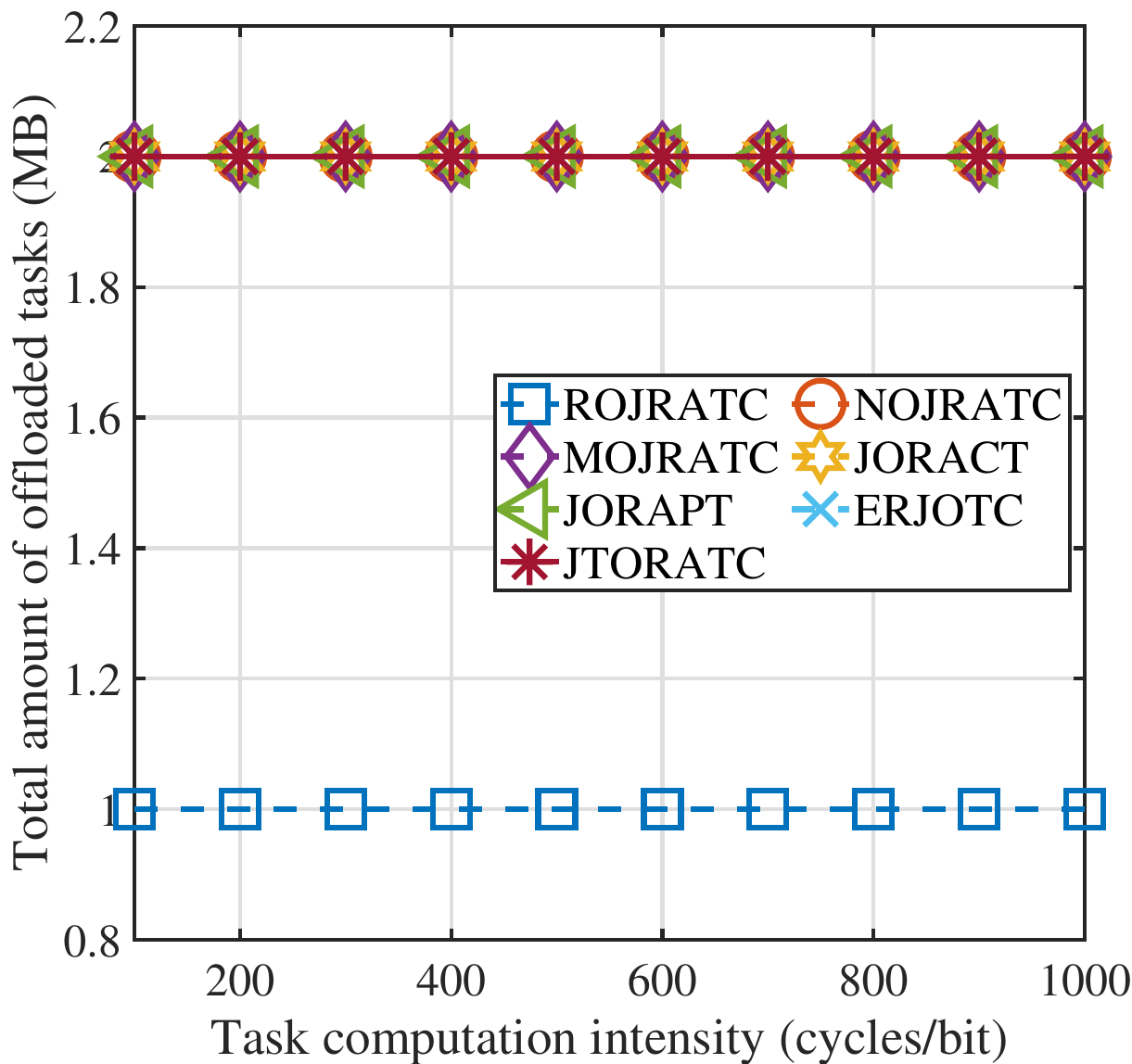}
		\end{minipage}
	}
	\centering
	\caption{System performance with task computation intensity. (a) Objective function value. (b) Total task completion delay. (c) Total UAV energy consumption. (d) Total amount of offloaded tasks.}
	\label{fig_computation_intensity}
	\vspace{-0.7em}
\end{figure*}

\textbf{\textit(2) {Impact of Task Computation Intensity.}} Figs. \ref{fig_computation_intensity}(a), \ref{fig_computation_intensity}(b), \ref{fig_computation_intensity}(c), and \ref{fig_computation_intensity}(d) reveal the effects of task computation intensity on the objective function value, total task completion delay, total UAV energy consumption, and total amount of offloaded tasks. As can be seen, as the computational intensity increases, the seven approaches exhibit similar upward trends in terms of objective function value, total task completion delay and total UAV energy consumption. This is because higher computational intensity indicates heavier system workloads, which further leads to higher computational requirements. Consequently, more frequent task offloading, computation resource allocation and adjustments of UAV trajectory are necessary, which potentially results in additional overhead on computation and energy consumption.

\par Moreover, it can be seen from Fig. \ref{fig_computation_intensity}(a) that the proposed JTORATC outperforms other approaches in terms of objective function value. This is because JTORATC iteratively optimizes the decisions of task offloading, computation resource allocation, and UAV trajectory control to adapt to scenarios with varying workloads. From Fig. \ref{fig_computation_intensity}(b), we can observe that with the increasing of task computation intensity, ROJRATC achieves a similar total completion delay to the proposed JTORATC by using the random offloading approach. However, the performance of ROJRATC in minimizing the objective function value and total UAV energy consumption is not as good as our proposed JTORATC approach. This is because the ROJRATC mainly relies on a random offloading strategy, which could not provide sufficient performance benefits in terms of the objective function, completion delay, and energy consumption, as the computational intensity of the tasks increases. In contrast, the proposed JTORATC takes a more comprehensive approach for system optimization. 

\par From Fig. \ref{fig_computation_intensity}(c), it can be observed that ERJOTC shows optimal performance in terms of total UAV energy consumption. This is because ERJOTC helps to ensure the load balance of each UAV during task execution, reduce the excessive use of computing resources, and achieve an even distribution of overall computing energy consumption. However, the performance of ERJOTC is slightly inferior to our proposed JTORATC approach in objective function value and total task completion delay aspects. While from Fig. \ref{fig_computation_intensity}(d), we can see that for the total amount of offloaded tasks, the seven approaches all show a constant trend regardless of the varying task computation intensity. Similarly, because the total amount of tasks completed by the UAV depends on the amount of tasks generated by the users, and has nothing to do with the task computation intensity.


\par In conclusion, the simulation results in Fig. \ref{fig_computation_intensity} indicate the superiority of the proposed JTORATC in both light and heavy computational scenarios.

\begin{figure*}[!hbt] 
	\centering
	\subfigure[]
	{
		\begin{minipage}[t]{0.23\linewidth}
			\centering
		\includegraphics[width=1.64in]{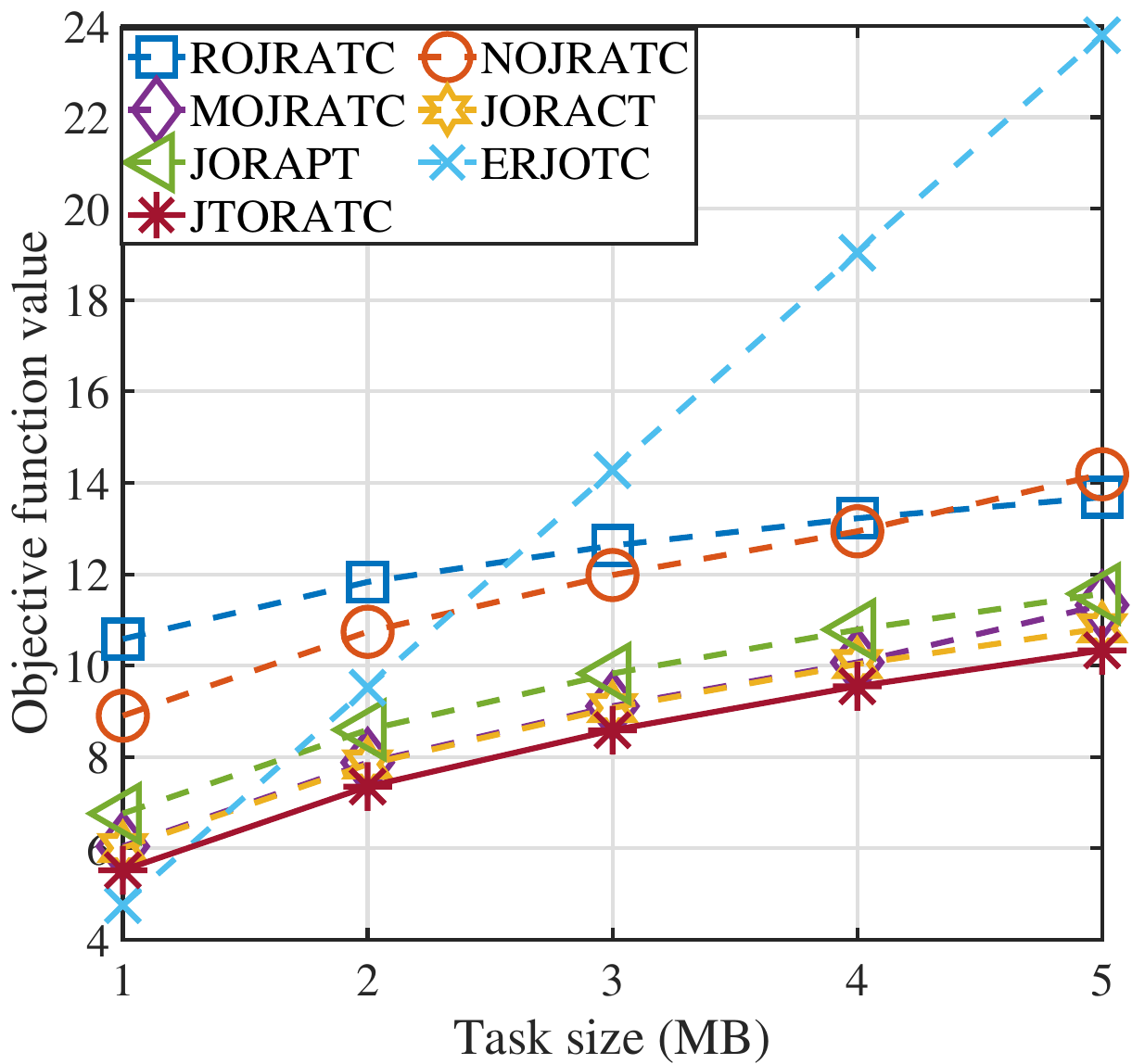}
		\end{minipage}
	}
	\subfigure[]
	{
		\begin{minipage}[t]{0.23\linewidth}
			\centering
		\includegraphics[width=1.64in]{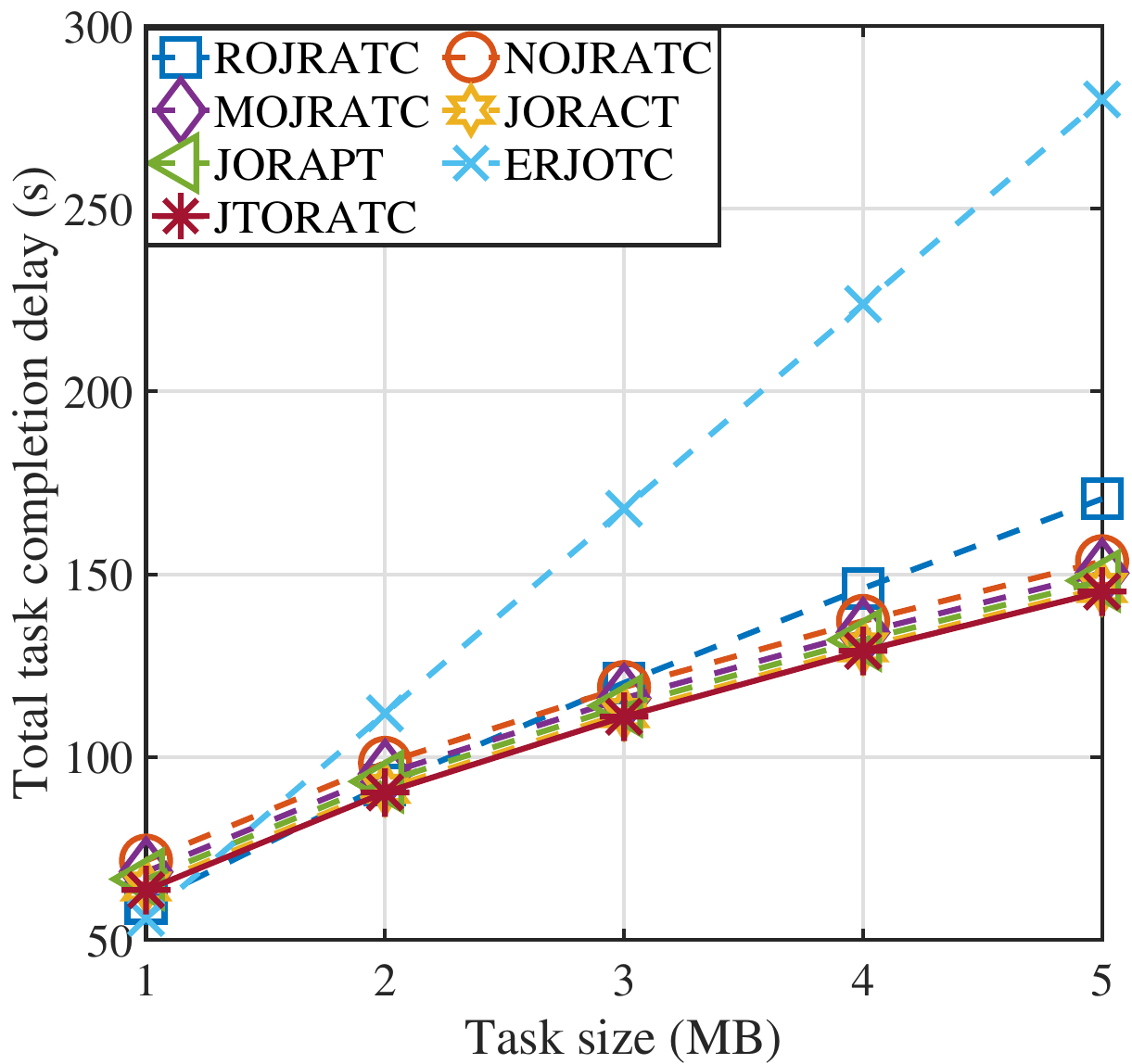}	
		\end{minipage}
	}
	\subfigure[]
	{
		\begin{minipage}[t]{0.23\linewidth}
			\centering
		\includegraphics[width=1.64in]{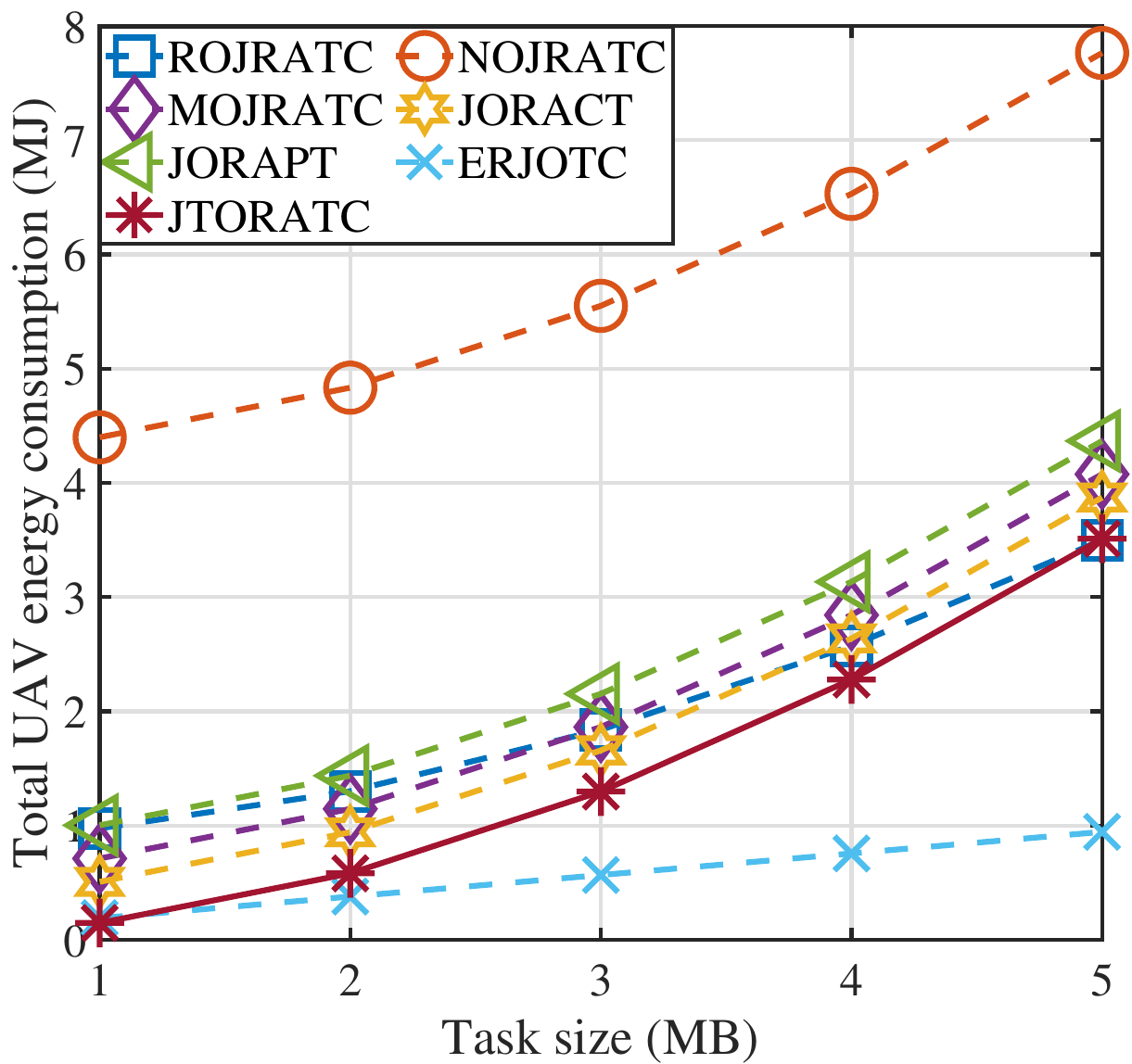}
		\end{minipage}
	}
        \subfigure[]
	{
		\begin{minipage}[t]{0.23\linewidth}
			\centering
		\includegraphics[width=1.64in]{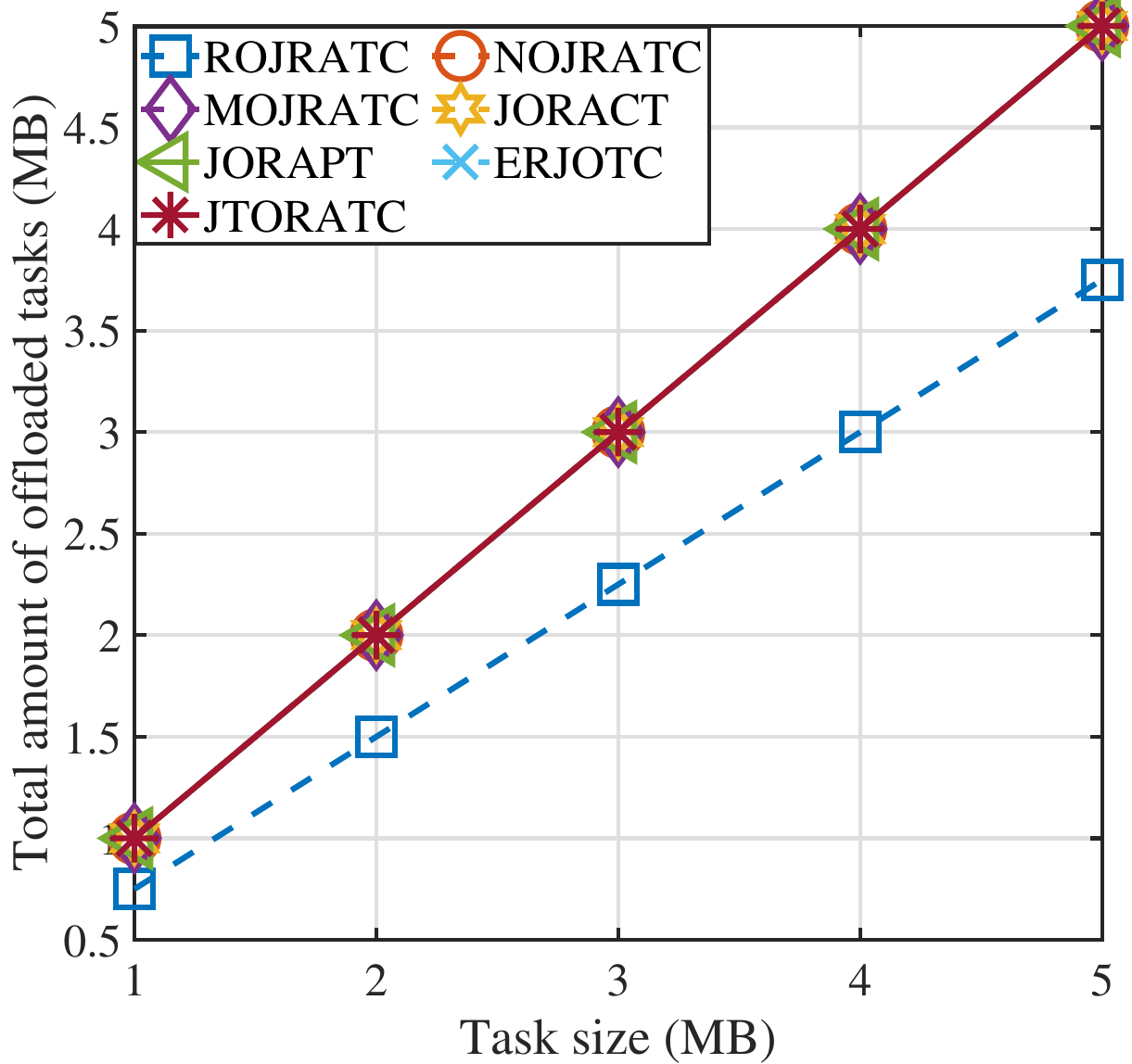}
		\end{minipage}
	}
	\centering
	\caption{System performance with task size. (a) Objective function value. (b) Total task completion delay. (c) Total UAV energy consumption. (d) Total amount of offloaded tasks.}
	\label{fig_task_size}
	\vspace{-0.7em}
\end{figure*}

\textbf{\textit(3) {Impact of Task Size.}} Figs. \ref{fig_task_size}(a), \ref{fig_task_size}(b), \ref{fig_task_size}(c), and \ref{fig_task_size}(d) depict the effects of task size on the objective function value, the total task completion delay, the total UAV energy consumption, and the total amount of offloaded tasks. It can be seen from Fig. \ref{fig_task_size} that the seven approaches exhibit upward trends as the task size increases. This is because the increasing task size indicates heavier workloads, which could further lead to higher costs of processing delay and energy consumption.

\par From Figs. \ref{fig_task_size}(a), \ref{fig_task_size}(b), and \ref{fig_task_size}(c), it can be observed that when the task size is small (less than 1.5MB), the performance of ERJOTC is better than the proposed JTORATC approach in objective function and total task completion delay aspects. However, when the task size is large (more than 3MB), ERJOTC has the worst performance of all approaches. Moreover, for the total UAV energy consumption, ERJOTC shows the best performance. This is because when the task size is small, resources are allocated uniformly, then ERJOTC may be more efficient. However, when the task size is large, ERJOTC does not fully consider the complexity and different requirements of the task, which may lead to performance degradation. In addition, the average allocation strategy of ERJOTC ensures a relatively uniform distribution of tasks among UAVs and avoids excessive loads, thus reducing the energy consumption of UAVs.

\par Moreover, it can be seen from Fig. \ref{fig_task_size}(d) that the seven approaches all show a upward trend as task size increases for the total amount of offloaded tasks, because the total amount of tasks generated by users is increasing, and the characteristics and requirements of the tasks may also change.

For the total amount of offloaded tasks, the seven approaches all show a upward trend as task size increases since the total amount of tasks generated by users is increasing, and the characteristics and requirements of the tasks may also change.

\par Accordingly, the simulation results in Fig. \ref{fig_task_size} demonstrate the superiority of the proposed JTORATC to adapt to heavy-loaded scenarios.

\begin{figure*}[!hbt] 
	\centering
	\subfigure[]
	{
		\begin{minipage}[t]{0.23\linewidth}
			\centering
		\includegraphics[width=1.64in]{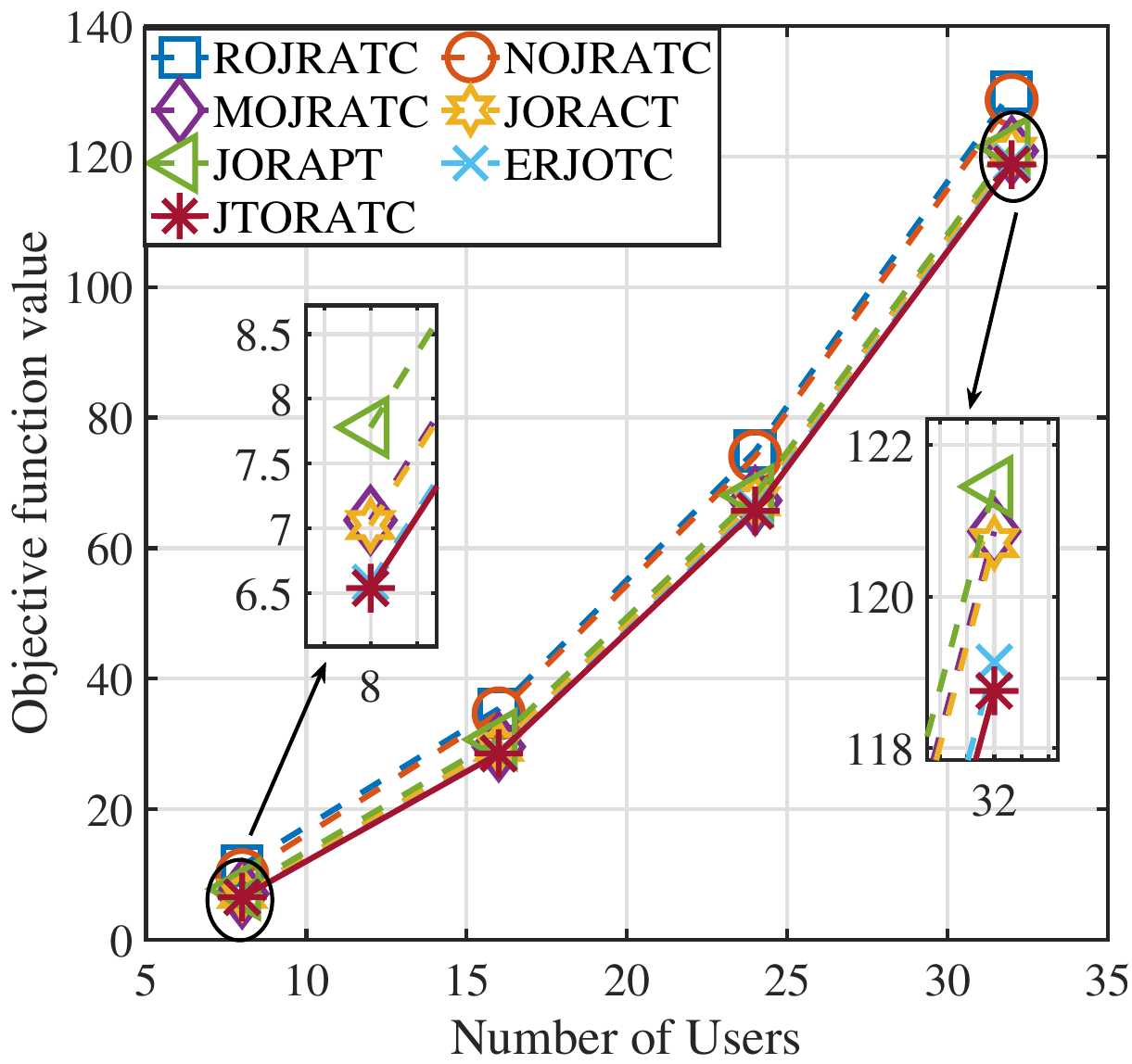}
		\end{minipage}
	}
	\subfigure[]
	{
		\begin{minipage}[t]{0.23\linewidth}
			\centering
		\includegraphics[width=1.64in]{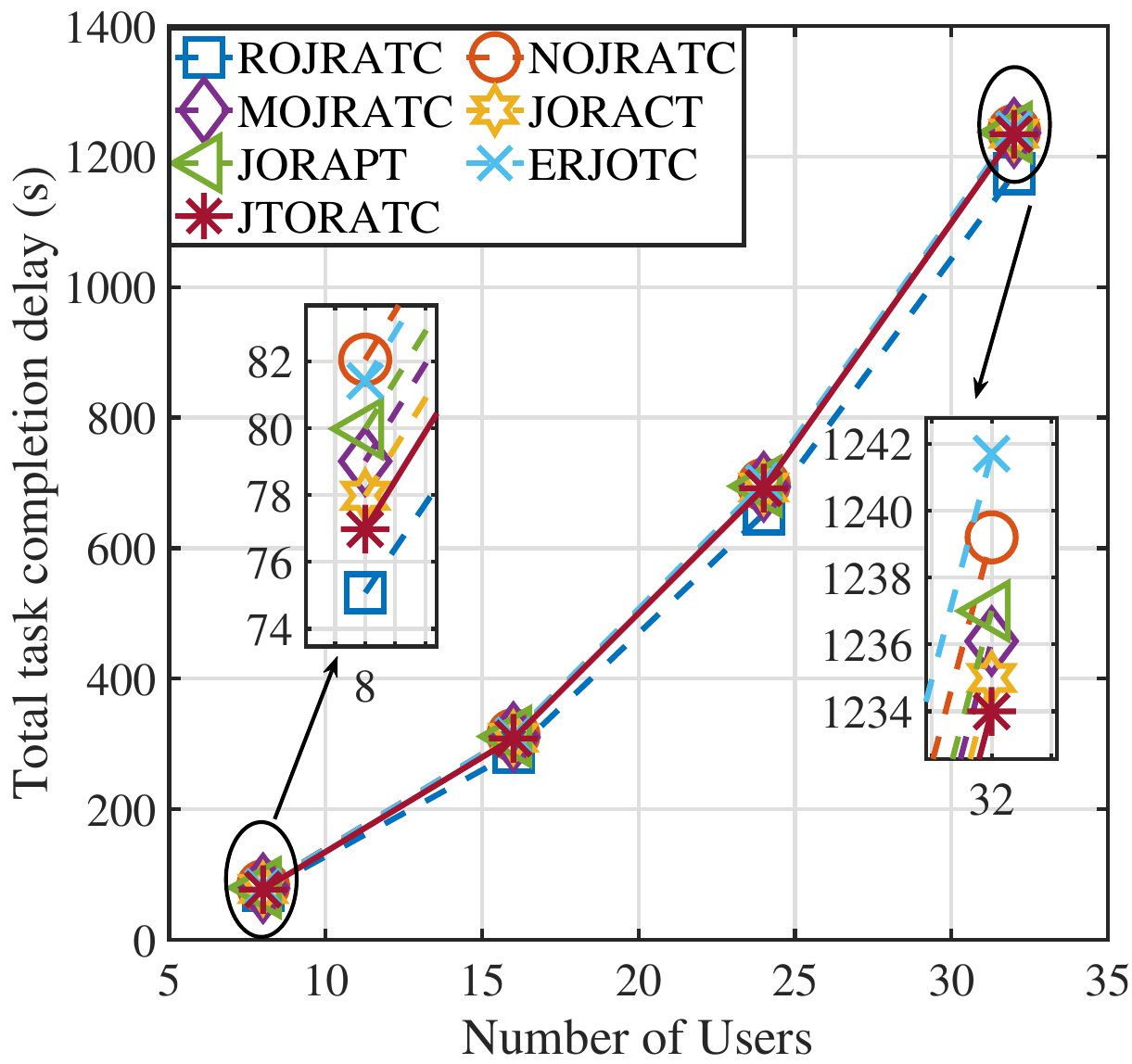}	
		\end{minipage}
	}
	\subfigure[]
	{
		\begin{minipage}[t]{0.23\linewidth}
			\centering
		\includegraphics[width=1.64in]{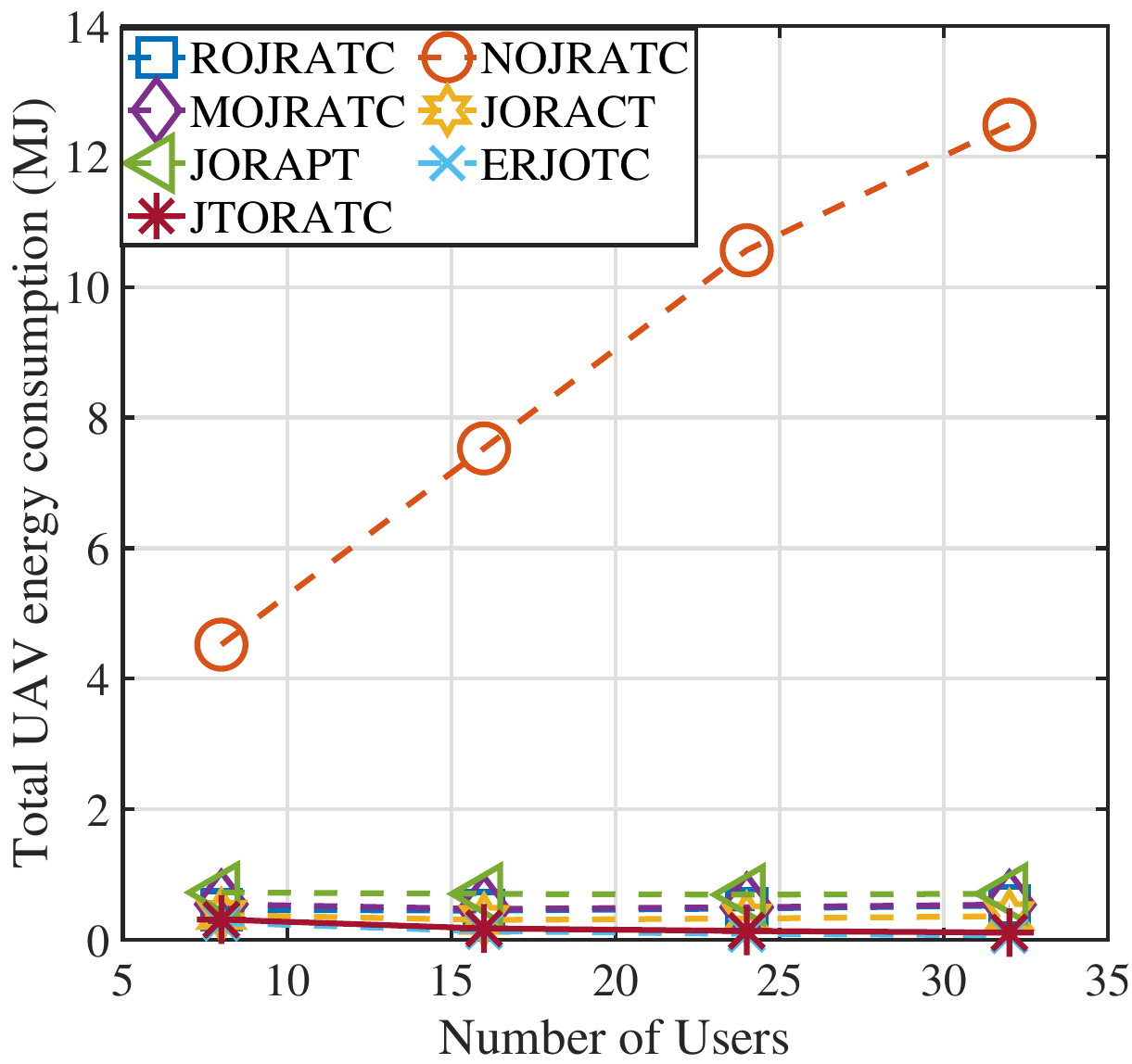}
		\end{minipage}
	}
        \subfigure[]
	{
		\begin{minipage}[t]{0.23\linewidth}
			\centering
		\includegraphics[width=1.64in]{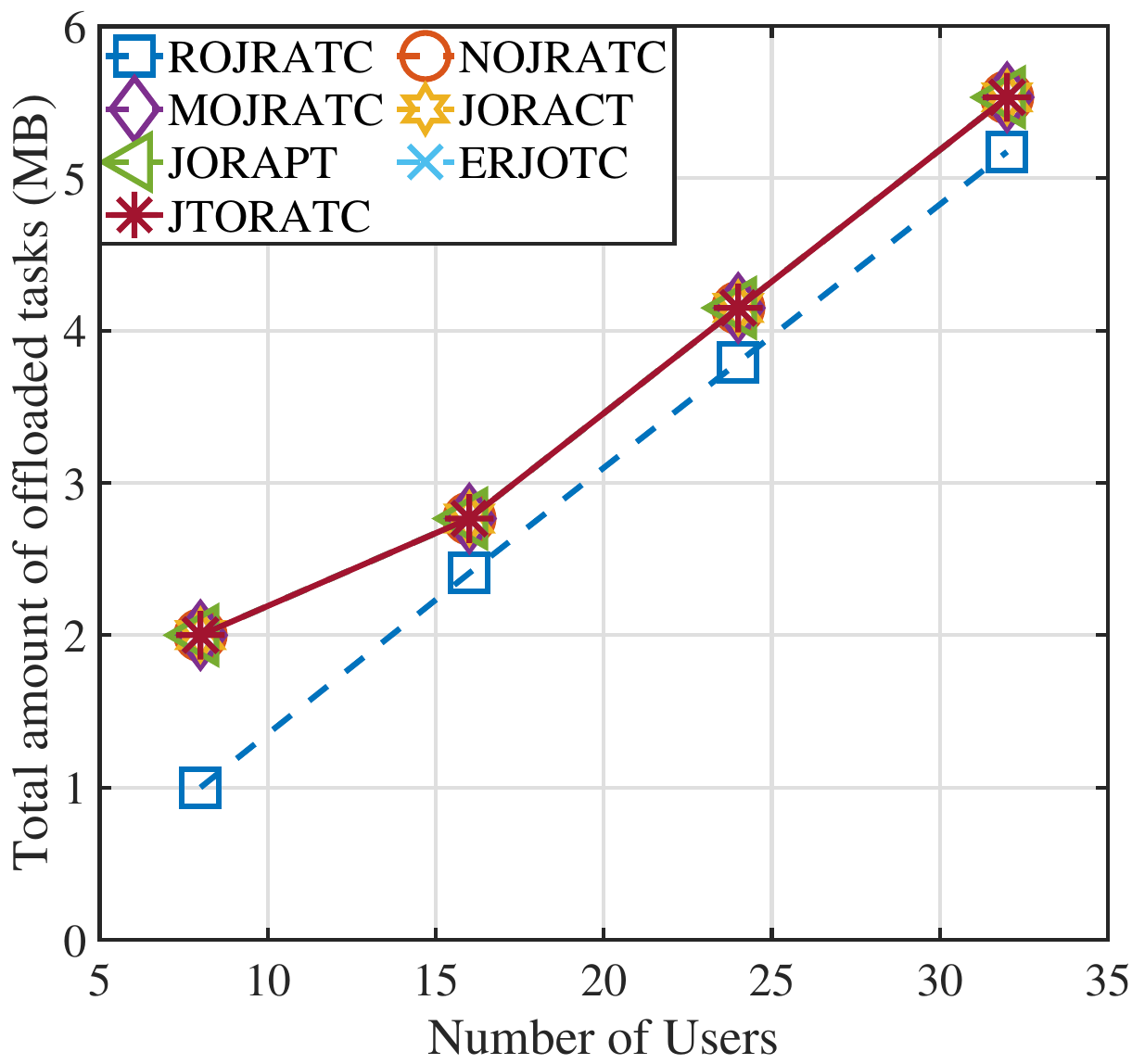}
		\end{minipage}
	}
	\centering
	\caption{System performance with number of users. (a) Objective function value. (b) Total task completion delay. (c) Total UAV energy consumption. (d) Total amount of offloaded tasks.}
	\label{fig_number_of_user}
	\vspace{-0.7em}
\end{figure*}

\textbf{\textit(4) {Impact of Number of Users.}} Figs. \ref{fig_number_of_user}(a), \ref{fig_number_of_user}(b), \ref{fig_number_of_user}(c), and \ref{fig_number_of_user}(d) present that with the increasing of the number of users, the objective function value, the total task completion delay, the total UAV energy consumption, and the total amount of offloaded tasks of the seven approaches gradually increase. The reason is that as the number of users increases, the total number of tasks that need to be handled in the system also increases. This leads to a surge in resource competition and an increase in the load of communication networks. 

\par Fig. \ref{fig_number_of_user}(a) illustrates that our proposed JTORATC approach achieves the best performance in minimizing the value of the objective function. This can be attributed to the fact that our proposed JTORATC successfully balances the system parameters by employing iterative optimization, and thus obtains a significant advantage in the overall performance.

\par From Figs. \ref{fig_number_of_user}(b), \ref{fig_number_of_user}(c), and \ref{fig_number_of_user}(d), it can be observed that ROJRATC performs best in terms of total task completion delay. The random nature of ROJRATC can lead to tasks being assigned and completed more quickly. However, it should be noted that ROJRATC performs relatively poorly in terms of total UAV energy consumption and total amount of offloaded tasks. This suggests that we must fully balance energy consumption and task completion efficiency while considering delay. In contrast, our proposed JTORATC approach takes this balance into account. Moreover, the ERJOTC has a good performance in reducing the total UAV energy consumption, mainly because evenly allocating resources avoids some resources being idle or overloaded. At the same time, because of the uniform allocation of resources, there is a certain sacrifice in the task completion delay, which is far less than our proposed JTORATC approach. This may be because task offloading is not flexible enough or resources cannot be optimized for the specific needs of the task.

\par Consequently, this set of simulation results highlights the importance of our proposed JTORATC approach in terms of scalability and robustness of the system.

\section{Conclusion}
\label{sec_conclusion}

\par In this work, we study the multi-objective optimization of task offloading, computation resource
allocation and UAV trajectory control in multi-UAV-assisted MEC systems. We first formulate the multi-objective optimization problem with the aim of minimizing the total task completion delay, reducing the total UAV energy consumption, and maximizing the total amount of offloaded tasks. To solve the problem, we transform the multi-objective optimization problem into a single-objective optimization problem and then propose the JTORATC to solve the transformed problem at a lower cost. Simulation results demonstrate that the proposed JTORATC has superior performance in terms of the total task completion delay and total UAV energy consumption. Specifically, JTORATC demonstrates better adaptability to the scenarios with varying computing capacities of UAVs, and it shows superior scalability in both light and heavy workload scenarios.

\ifCLASSOPTIONcaptionsoff
\newpage
\fi

\bibliographystyle{IEEEtran}
\bibliography{references.bib}
\vspace{-23pt}
\begin{IEEEbiography}
[{\includegraphics[width=1in,height=1.25in,clip,keepaspectratio]
{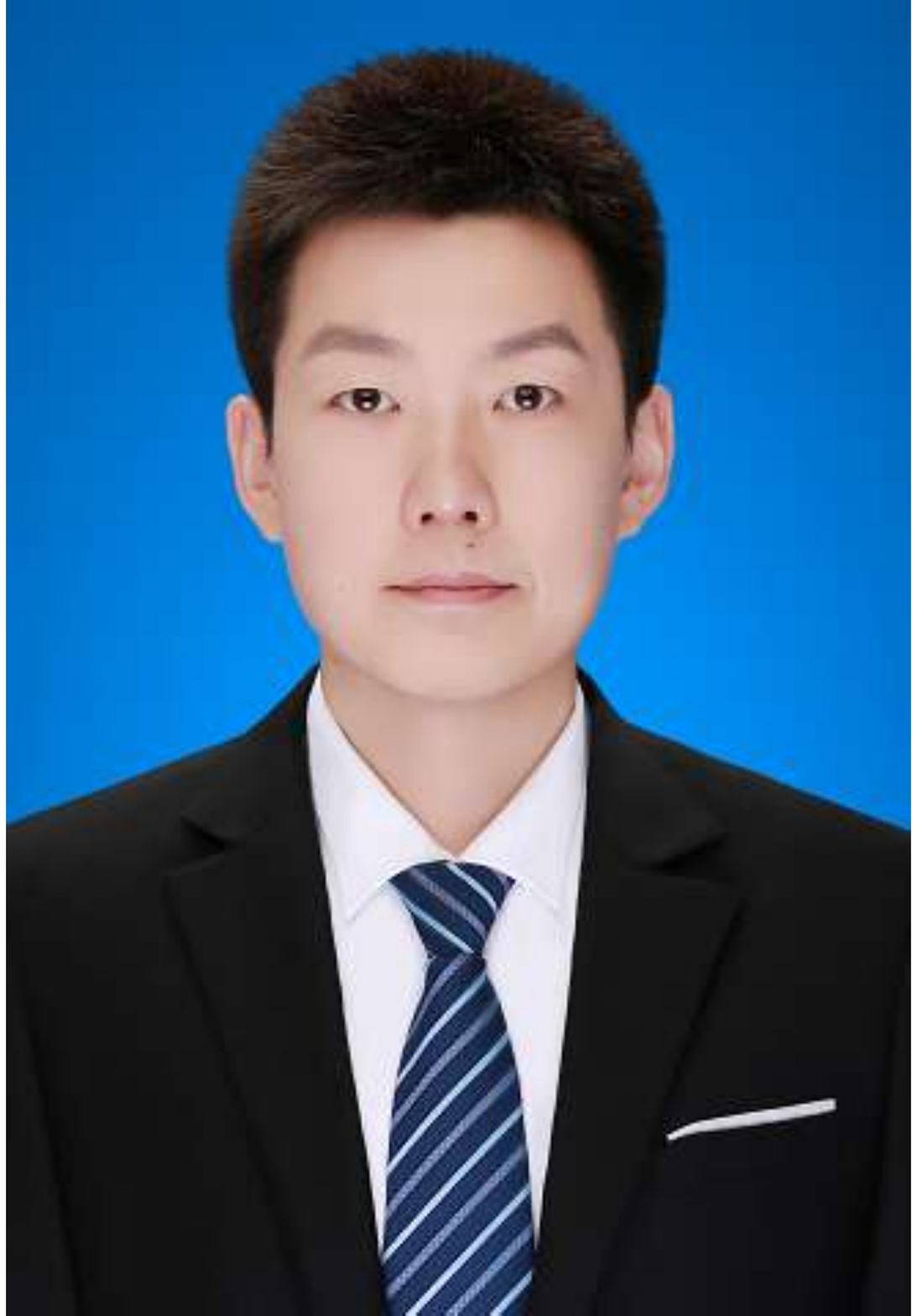}}]{Geng Sun} (S'17-M'19) received the B.S. degree in communication engineering from Dalian Polytechnic University, and the Ph.D. degree in computer science and technology from Jilin University, in 2011 and 2018, respectively. He was a Visiting Researcher with the School of Electrical and Computer Engineering, Georgia Institute of Technology, USA. He is an Associate Professor in College of Computer Science and Technology at Jilin University, and his research interests include wireless networks, UAV communications, collaborative beamforming and optimizations.
\end{IEEEbiography}

\begin{IEEEbiography}[{\includegraphics[width=1in,height=1.25in,clip,keepaspectratio]{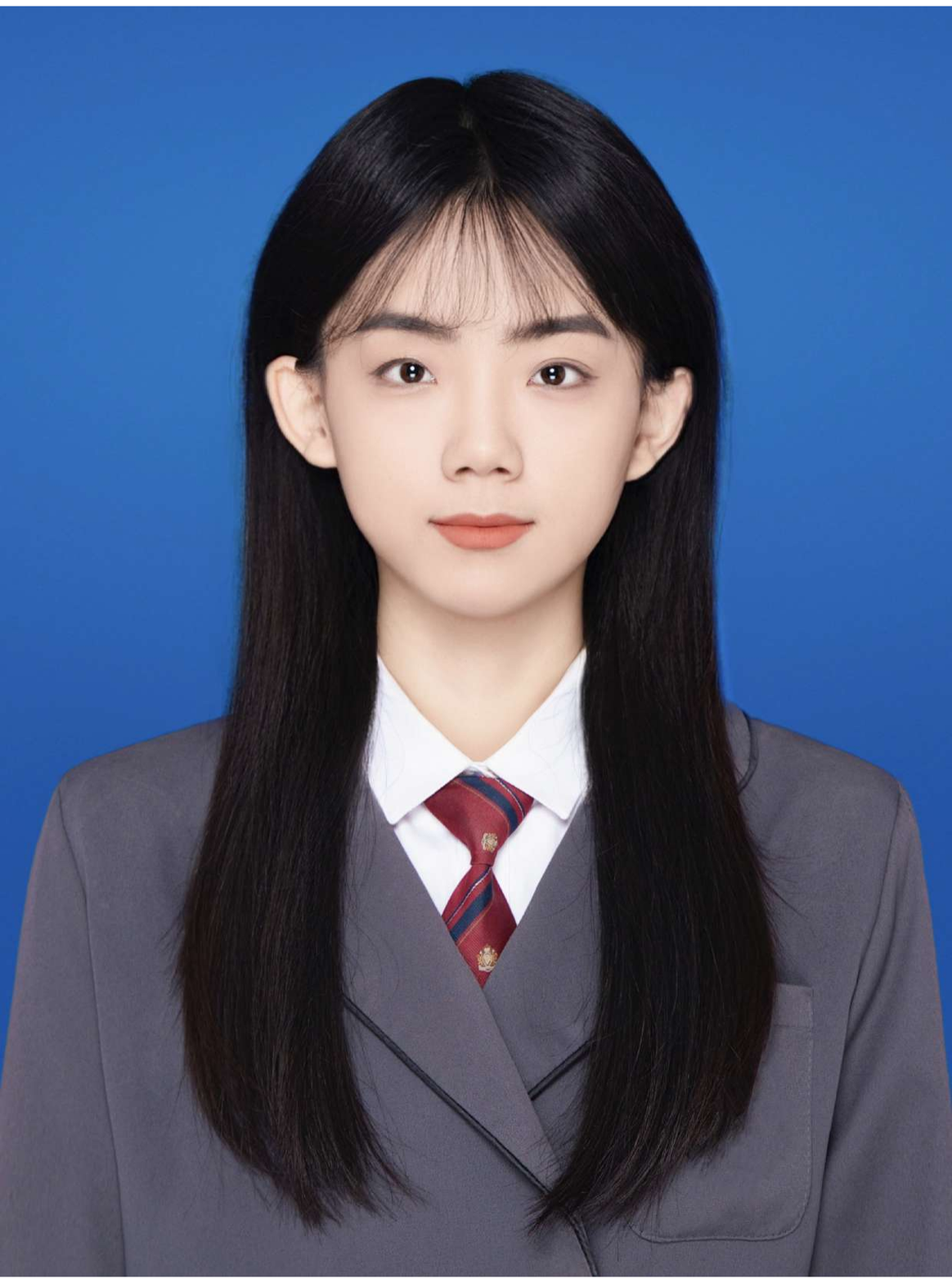}}]{Yixian Wang} received a BS degree in Software engineering from Shanxi University, Shanxi, China, in 2023. She is currently working toward the MS degree in Computer Science and Technology at Jilin University, Changchun, China. Her research interests include vehicular networks and edge computing.
\end{IEEEbiography}

\begin{IEEEbiography}[{\includegraphics[width=1in,height=1.25in,clip,keepaspectratio]
{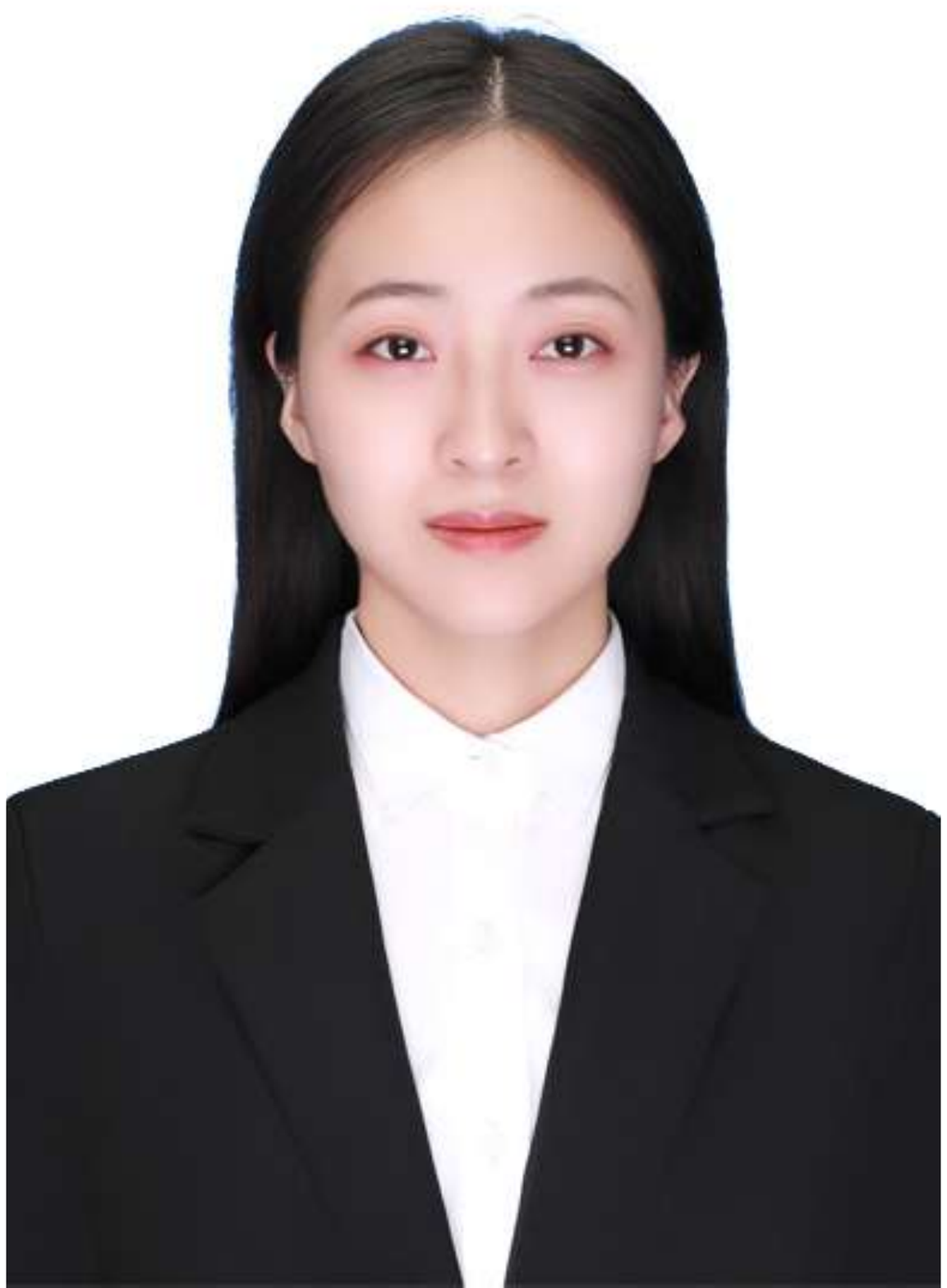}}]{Zemin Sun} received a BS degree in Software Engineering, an MS degree and a Ph.D degree in Computer Science and Technology from Jilin University, Changchun, China, in 2015, 2018, and 2022, respectively. Her research interests include vehicular networks, edge computing, and game theory. 
\end{IEEEbiography}

\begin{IEEEbiography}[{\includegraphics[width=1in,height=1.25in,clip,keepaspectratio]{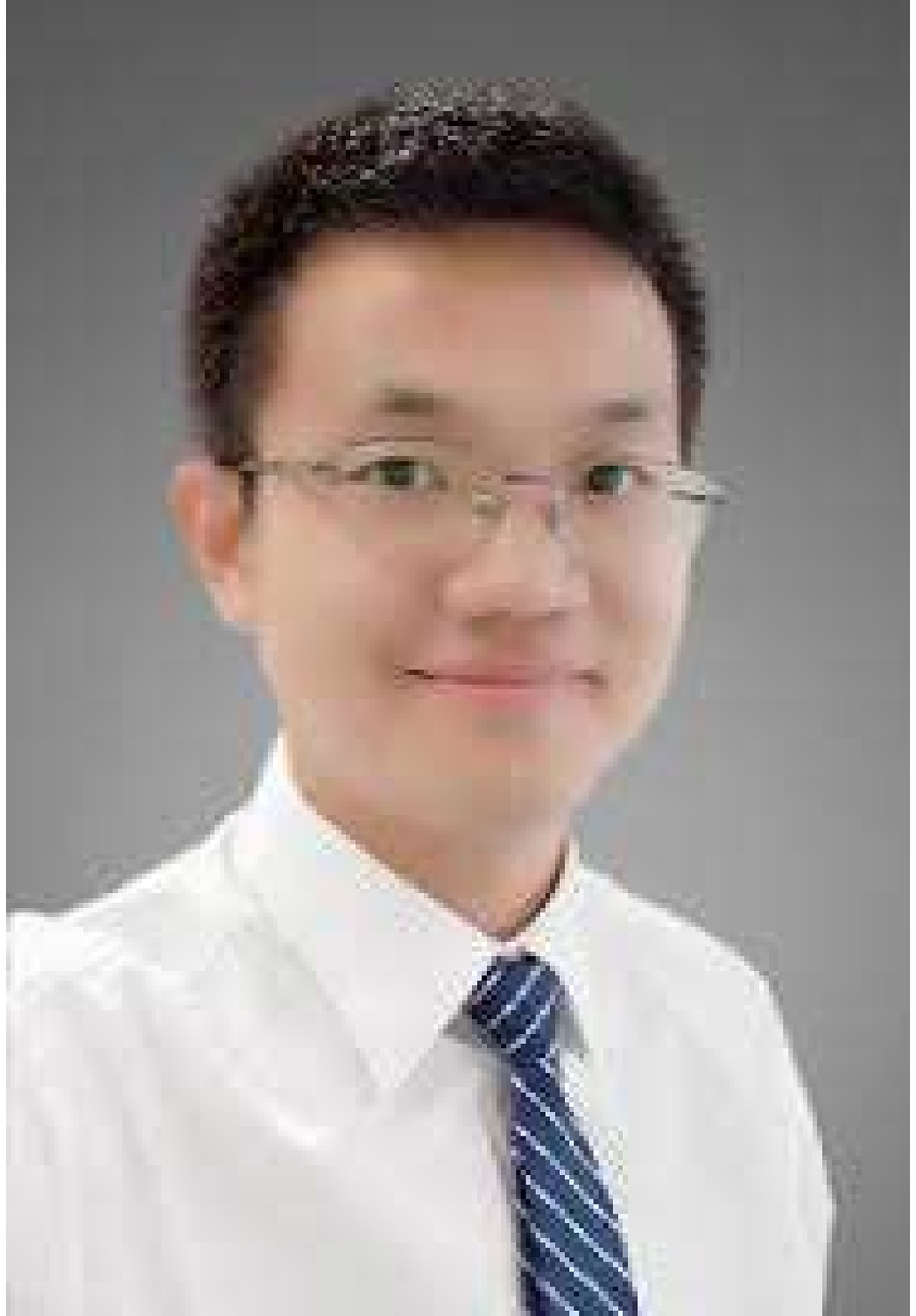}}]{Qingqing Wu} (S’13-M’16-SM’21) received the B.Eng. and the Ph.D. degrees in Electronic Engineering from South China University of Technology and Shanghai Jiao Tong University (SJTU) in 2012 and 2016, respectively. From 2016 to 2020, he was a Research Fellow in the Department of Electrical and Computer Engineering at National University of Singapore. He is currently an Associate Professor with Shanghai Jiao Tong University. His current research interest includes intelligent reflecting surface (IRS), unmanned aerial vehicle (UAV) communications, and MIMO transceiver design. He has coauthored more than 100 IEEE journal papers with 26 ESI highly cited papers and 8 ESI hot papers, which have received more than 18,000 Google citations. He was listed as the Clarivate ESI Highly Cited Researcher in 2022 and 2021, the Most Influential Scholar Award in AI-2000 by Aminer in 2021 and World’s Top 2\% Scientist by Stanford University in 2020 and 2021.
\end{IEEEbiography}

\begin{IEEEbiography}[{\includegraphics[width=1in,height=1.25in,clip,keepaspectratio]{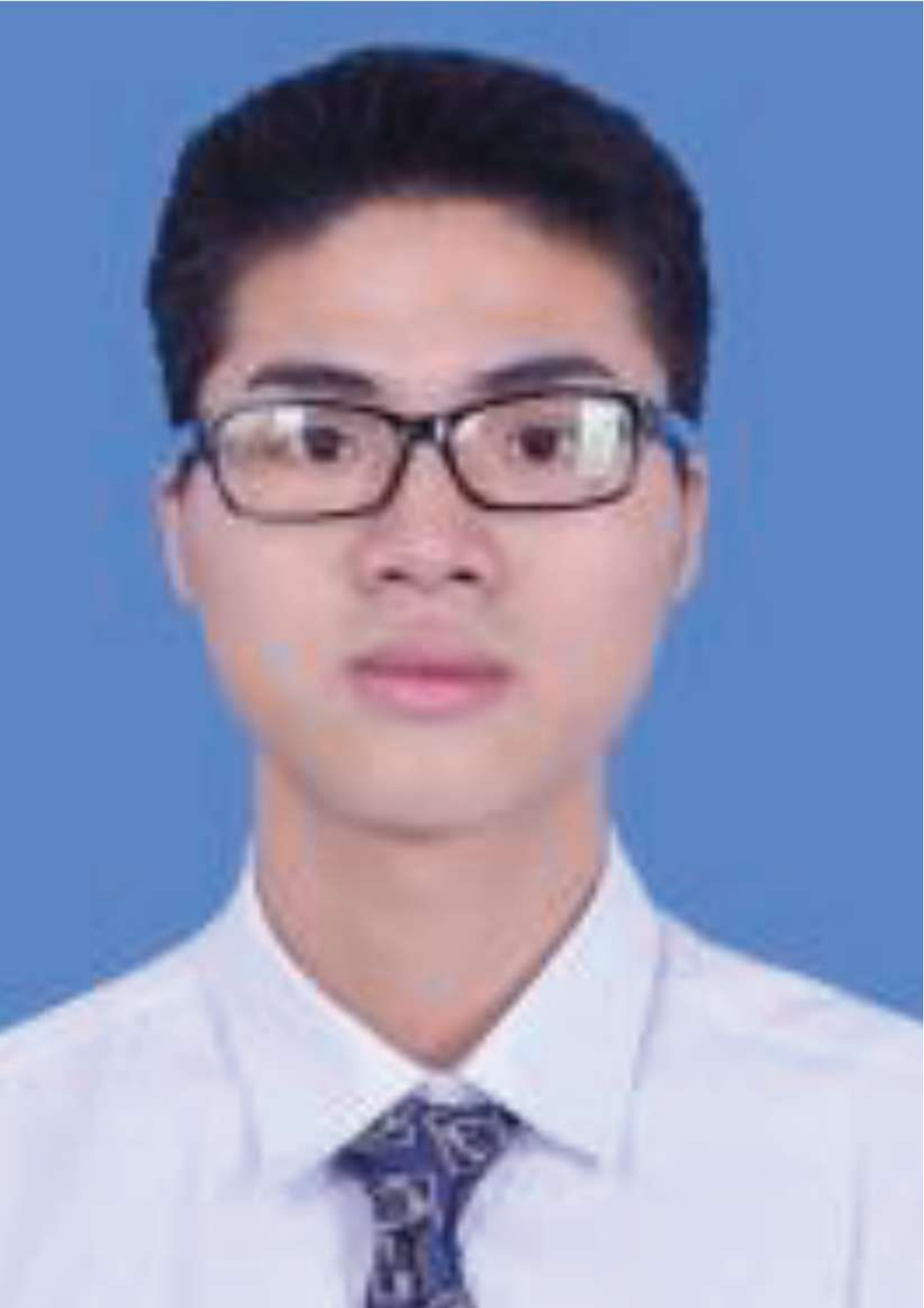}}]{Jiawen Kang} received the M.S. degree and the Ph.D. degree from the Guangdong University of Technology, China, in 2015 and 2018. He is currently a full professor at the Guangdong University of Technology. He was a postdoc at Nanyang Technological University from 2018 to 2021, Singapore. His research interests mainly focus on blockchain, security, and privacy protection in wireless communications and networking. 
\end{IEEEbiography}

\begin{IEEEbiography}[{\includegraphics[width=1in,height=1.25in,clip,keepaspectratio]{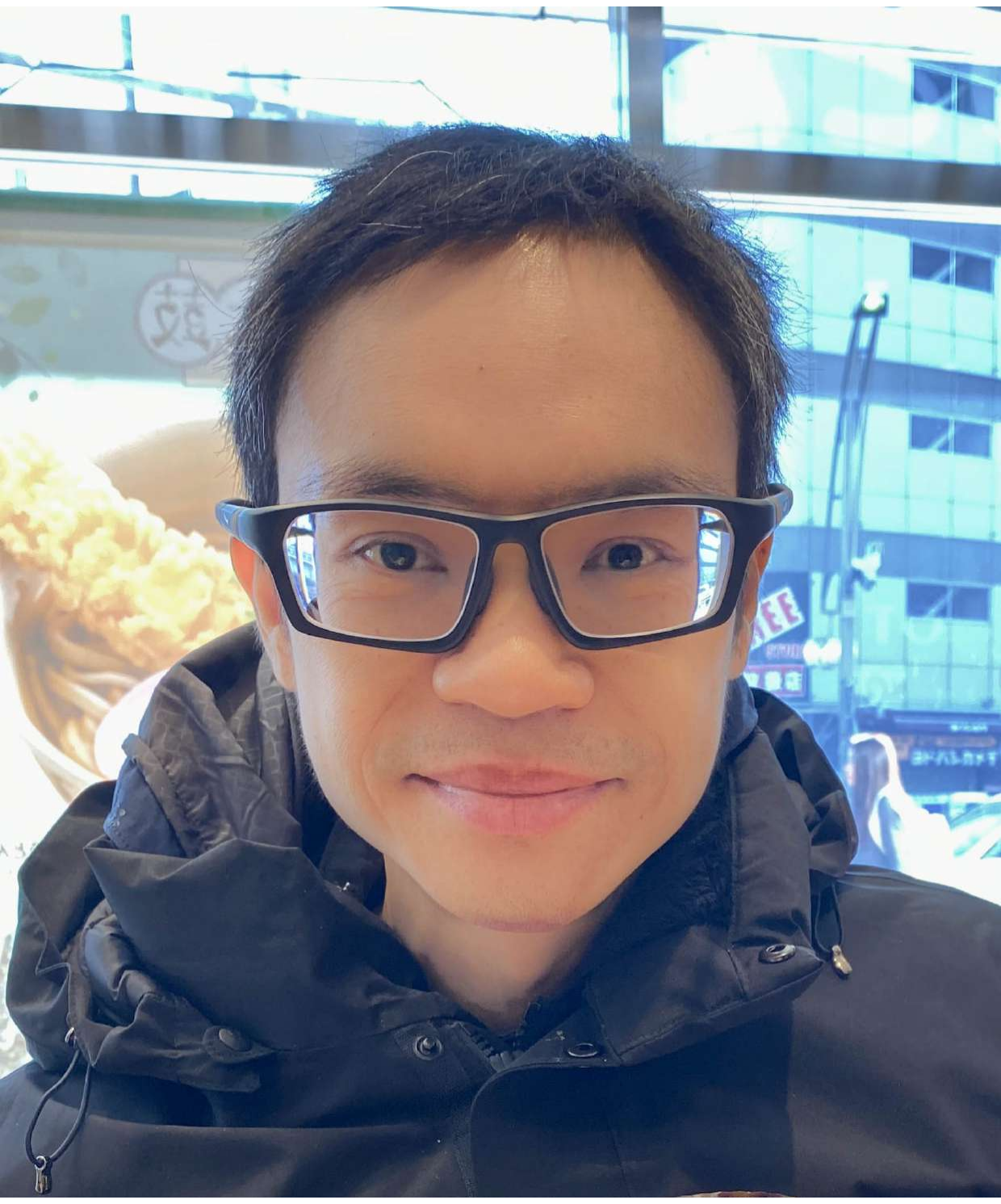}}]{Dusit Niyato} (Fellow, IEEE) received the B.Eng. degree from the King Mongkuts Institute of Technology Ladkrabang (KMITL), Thailand, in 1999, and the Ph.D. degree in electrical and computer engineering from the University of Manitoba, Canada, in 2008. He is currently a Professor with the School of Computer Science and Engineering, Nanyang Technological University, Singapore. His research interests include the Internet of Things (IoT), machine learning, and incentive mechanism design.
\end{IEEEbiography}

\begin{IEEEbiography}[{\includegraphics[width=1in,height=1.25in,clip,keepaspectratio]{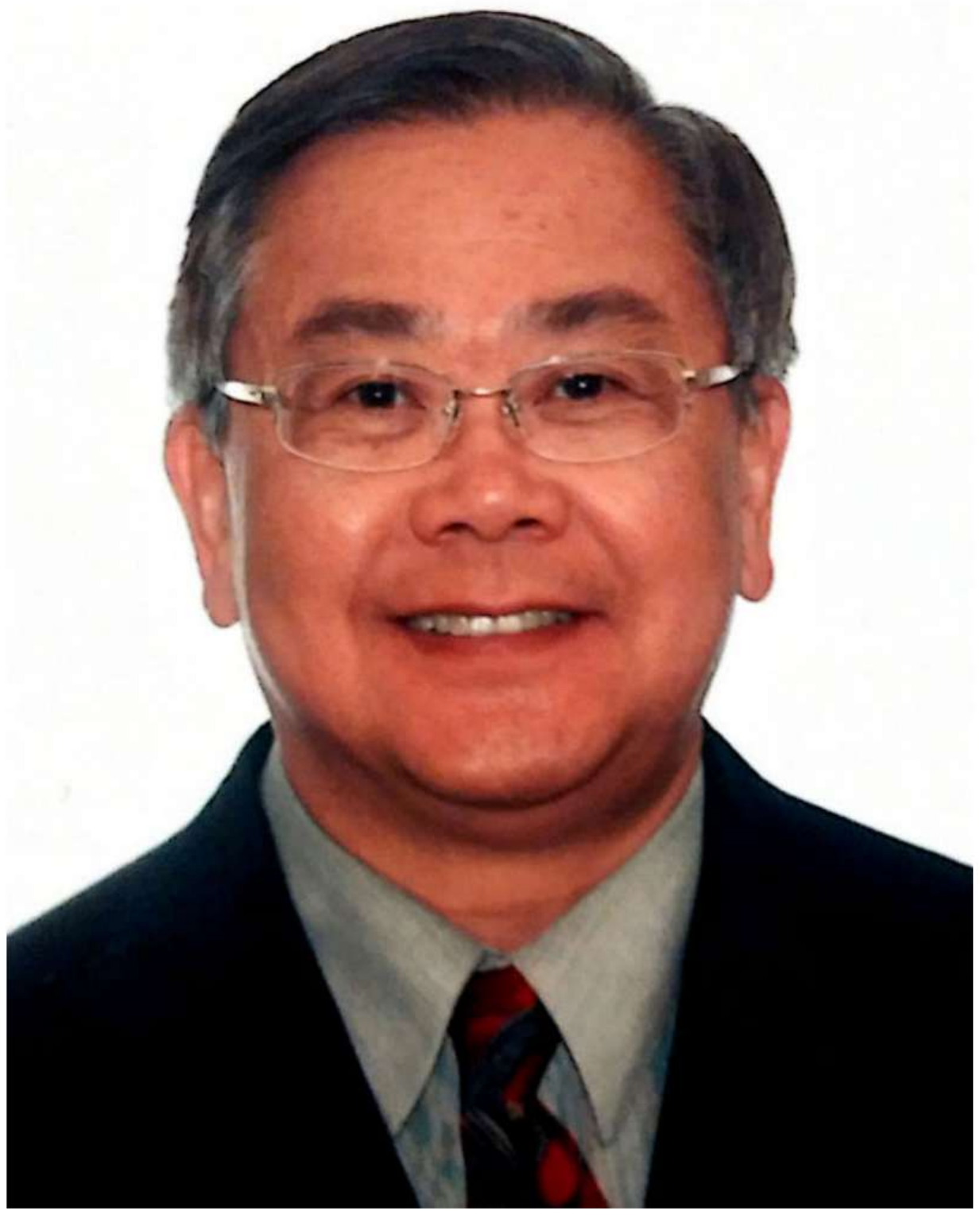}}]{Victor C. M. Leung} (Life Fellow, IEEE) is a Distinguished Professor of computer science and software engineering with Shenzhen University,
China. He is also an Emeritus Professor of electrial and computer engineering and the Director of the Laboratory for Wireless Networks and Mobile Systems at the University of British Columbia (UBC). His research is in the broad areas of wireless networks and mobile systems. He has co-authored more than 1300 journal/conference papers and book chapters. Dr. Leung is serving on the editorial boards of IEEE Transactions on Green Communications and Networking, IEEE Transactions on Cloud Computing, IEEE Access, and several other journals. He received the IEEE Vancouver Section Centennial Award, 2011 UBC Killam Research Prize, 2017 Canadian Award for Telecommunications Research, and 2018 IEEE TCGCC Distinguished Technical Achievement Recognition Award. He co-authored papers that won the 2017 IEEE ComSoc Fred W. Ellersick Prize, 2017 IEEE Systems Journal Best Paper Award, 2018 IEEE CSIM Best Journal Paper Award, and 2019 IEEE TCGCC Best Journal Paper Award. He is a Life Fellow of IEEE, and a Fellow of the Royal Society of Canada, Canadian Academy of Engineering, and Engineering Institute of Canada. He is named in the current Clarivate Analytics list of Highly Cited Researchers.
\end{IEEEbiography}

\end{document}